\documentclass[preprint,showpacs,showkeys,preprintnumbers,amsmath,amssymb,aps,groupedaddress]{revtex4}
\usepackage{graphicx}       
\usepackage{psfrag}
\usepackage{tikz}
\usepackage{mathptmx}
\usetikzlibrary{arrows,shapes,backgrounds}
\usepackage{ifthen}
\newboolean{includefigs}
\setboolean{includefigs}{true}
\newboolean{psfigs}
\setboolean{psfigs}{false}
\newboolean{tikzfigs}
\setboolean{tikzfigs}{true}

\newcommand{\condcomment}[2]{\ifthenelse{#1}{#2}{}}

\newcommand{\diag}{\mathop{\mathrm{diag}}}
\newcommand{\constant}{\mathop{\mathrm{constant}}}
\newcommand{\comment}[1]{}

\begin{document}


\title{An anisotropic, non-singular early universe model leading to a realistic cosmology}



\author{Pierre-Philippe Dechant}
\email{p.dechant@mrao.cam.ac.uk}
\affiliation{Astrophysics Group, Cavendish Laboratory, J J Thomson Avenue, University of Cambridge, CB3 0HE, UK}

\author{Anthony N. Lasenby}
\email{a.n.lasenby@mrao.cam.ac.uk}
\affiliation{Astrophysics Group, Cavendish Laboratory, J J Thomson Avenue, University of Cambridge, CB3 0HE, UK}

\author{Michael P. Hobson}
\email{mph@mrao.cam.ac.uk}
\affiliation{Astrophysics Group, Cavendish Laboratory, J J Thomson Avenue, University of Cambridge, CB3 0HE, UK}


\date{\today}

\begin{abstract}
We present a novel cosmological model in which scalar field matter in
a biaxial Bianchi IX geometry leads to a non-singular `pancaking'
solution: the hypersurface volume goes to zero instantaneously at the
{`Big Bang'}, but all physical quantities, such as curvature invariants
and the matter energy density remain finite, and continue smoothly
through the Big Bang.
We demonstrate that there exist geodesics extending through the Big
Bang, but that there are also incomplete geodesics that
spiral infinitely around a topologically closed spatial dimension at
the Big Bang, rendering it, at worst, a quasi-regular singularity. The
model is thus reminiscent of the Taub-NUT vacuum solution in that it has  biaxial Bianchi IX geometry
and its evolution exhibits a dimensionality reduction at a quasi-regular
singularity; the two models are, however, rather different, as we will
show in a future work.  Here we concentrate on the cosmological
implications of our model and show how the scalar field drives both
isotropisation and inflation, thus raising the question of whether
structure on the largest scales was laid down at a time when the
universe was still oblate (as also suggested by
{\cite{PitrouUzan2007CosmoPerturbationsAnisotropic,
Pitrou2008PredictionsAnisotropicEra, Contaldi2007InflationaryPertubationsInBianchi}}).  We also
discuss the stability of our model to small perturbations around
biaxiality and draw an analogy with cosmological perturbations. We
conclude by presenting a separate, bouncing solution, which
generalises the known bouncing solution in closed FRW universes.

\end{abstract}



\maketitle


\section{Introduction\label{int}}

At the core of most of theoretical cosmology lie the assumptions of
homogeneity and isotropy. These were originally motivated by the
cosmological principle and the mathematical tractability of the
resulting FRW models. However, the observed universe is obviously
neither homogeneous nor isotropic, so these symmetries can only ever
be approximate. Thus the question arises as to which assumptions we
can relax whilst maintaining analytical tractability. We choose here
to consider more general, anisotropic cosmologies, whilst keeping the
assumption of (large-scale) spatial homogeneity. The resulting class
of cosmologies is collectively known as the Bianchi models.

Scalar fields are ubiquitous in theories of high energy physics. The
Standard Model of particle physics postulates the scalar Higgs
particle, and Superstring Theory and string-inspired models motivate a
plethora of moduli fields from compactifications, dilatons, radions
etc. Scalar fields are also of interest in cosmology, as they
can drive periods of accelerated expansion of the universe in a
relatively straightforward and plausible manner. Cosmological scalar
fields could therefore account for the posited period of inflation and
the apparent present period of $\Lambda$-domination.

It is therefore natural to consider the effects of a scalar field
dominating the dynamics of a Bianchi model. However, the
phenomenological success of FRW models suggests that we should only
consider the Bianchi classes that allow FRW universes as special
cases. This narrows the phenomenologically interesting Bianchi types
down to $\mathrm{I}$, $\mathrm{V}$, $\mathrm{VII_{h}}$ and
$\mathrm{IX}$. From previous work
\cite{LasenbyDoran2005ClosedUniversesdSandInflation,
  LasenbyDoran2004ConformalModelsICsCMB} we are particularly
interested in closed models, which are still consistent with
observations \cite{Spergel2007WMAP3Cos}.  The only closed Bianchi type
that also allows for FRW subclasses is Bianchi IX, so we shall
consider the dynamics of a scalar field in a Bianchi IX model
\cite{Toporensky1999BianchiIXMassiveScalarField,
  Fay2005BianchiIXMassiveScalarFieldChaos, ColeyGoliath2000}. Bianchi
IX models are known to have very complicated dynamics, exhibiting
oscillatory singularities and chaos (mixmaster)
\cite{Misner1969Mixmaster,BelinskyKhalatnikovLifshitz1972,BelinskyKhalatnikovLifshitzKL1969,
  Ellis1969ClassOfHomogeneousCosmologicalModels,
  Novikov1973ProcessesNearSingularities,
  Ellis2006BianchiModelsThenAndNow, UgglaHeinzle2009mixmaster, UgglaHeinzle2009newproof}. In this light, it is even more
intriguing that the biaxial model we will consider is so well-behaved.

{It is interesting that our solution {provides} another example of a regular solution with
axial symmetry; for others, see Pitrou et al
\cite{PitrouUzan2007CosmoPerturbationsAnisotropic, Pitrou2008PredictionsAnisotropicEra} and Senovilla and collaborators \cite{Senovilla1990NewClassOfInhomogeneous,Senovilla1992GeneralClass,Senovilla1992Singularity-free,Fernandez2002AWideFamily}.
In the different context of inhomogeneous {but axially symmetric} cosmologies,
\cite{Senovilla1990NewClassOfInhomogeneous,Senovilla1992GeneralClass,Senovilla1992Singularity-free,Fernandez2002AWideFamily}
 also consider the important concept of geodesic completeness
of a spacetime.  Completeness is a criterion for the physical significance of a solution that is more stringent than mere geometric regularity, which
we will also need to address later.}

{Often inflation is invoked to justify the assumptions of (acausal) homogeneity and isotropy. However, this then raises
the question of how sensitive inflation itself is to the rather unnatural initial conditions of a homogeneous
initial state. For these considerations, the interested reader may refer to, for instance, \cite{GoldwirthPiran1992ICsforinflation}.
Here, however, we consider a Bianchi geometry as our starting assumption,  without invoking inflation in order
to justify it. For recent related literature concerning the case where a vector field
is present to drive a phase of anisotropic inflation, see for example \cite{Soda2008AnisotropicInflationVector, Contaldi2008Instability}.}

This paper is organised as follows. We begin with a brief introduction
to Bianchi models in Section \ref{cla}. In Section \ref{tri} we
construct a generic Bianchi IX model dominated by scalar field matter
and then specialise to the biaxial case in Section \ref{axi}.
({An alternative form of the Einstein field equations for this system,
 using the $3+1$ {covariant} approach, is presented in the appendix.})
We demonstrate scaling behaviour of solutions, and present a particular
series solution, in which one radius goes to zero at the Big Bang
(pancaking). We also consider geodesics through the Big Bang. We then
present a realistic cosmology based on our model in Section \ref{cos}
and, in Section \ref{pert}, the stability of such a model about
biaxiality. We finally present a separate, bouncing solution in
Section \ref{bounce}, before we conclude in Section \ref{con}.

\section{Bianchi Universes\label{cla}}

Bianchi universes are spatially homogeneous and therefore have a
3-dimensional group of isometries $G_{3}$ acting simply transitively
on spacelike hypersurfaces. The standard classification follows
Bianchi's (1897) classification of 3-parameter Lie groups
\cite{bianchi1897}.

We adopt the metric convention $\left(+ - - \,-\right)$. Roman letters
$a,b,c...$ from the beginning of the alphabet denote Lie algebra
indices.  Greek letters $\mu,\nu,\sigma...$ label spacetime indices,
whereas Roman letters $i,j,k...$ from the middle of the alphabet label
purely spatial ones.

The isometry group of a manifold is isomorphic to some Lie group $G$
and the Killing vectors obey
$\left[\xi_{\mu},\xi_{\nu}\right]=C_{\,\,\mu\nu}^{\sigma}\xi_{\sigma}$
where the $C_{\,\,\mu\nu}^{\sigma}$ are the structure constants of
$G$, so Lie groups can be used to describe symmetries, in particular,
isometries.

When studying symmetries, an invariant basis is often useful.  This is
a set of vector fields $X_{\mu}$ each of which is invariant under $G$,
i.e. has vanishing Lie derivative with respect to all the Killing
vectors such that
\begin{equation}
\left[\xi_{\mu},X_{\nu}\right]=0.\label{KVinv}
\end{equation}
Such a basis can be constructed simply by imposing this relation at a
point for some chosen set of independent vector fields and using the
Killing vectors to drag them out across the manifold. The
integrability condition for this set of first-order differential
equations amounts to demanding that the $C_{\,\,\mu\nu}^{\sigma}$ be
the structure constants of some group. The invariant vector fields in
fact satisfy
\begin{equation}
\left[X_{\mu},X_{\nu}\right]=-C_{\,\,\mu\nu}^{\sigma}X_{\sigma}.
\label{invbas}
\end{equation}
Denoting the duals of the $X_{\mu}$ by $\omega^{\mu}$, the
corresponding curl relations for the dual basis are
\begin{equation}
d\omega^{\mu}=\frac{1}{2}C_{\,\,\sigma\tau}^{\mu}\omega^{\sigma}\wedge\omega^{\tau}.
\label{curl}
\end{equation}
Because the $X_{\mu}$ are invariant vectors, the metric can now be
expressed as
\begin{equation}
ds^{2}=g_{\mu\nu}\omega^{\mu}\omega^{\nu},\label{constmet}
\end{equation}
for some $g_{\mu\nu}$.

Bianchi models can be constructed in various different ways.  They are
based on properties of a tetrad $\left\{ \mathbf{e}_{i}\right\} $ that
commutes with the basis of Killing vectors $\left\{ \xi_{j}\right\} $
which generate the symmetry group
\begin{equation}\label{tetradKV}
[\mathbf{e}_{i},\xi_{j}]=0.
\end{equation}

Here we have made use of the fact that for homogeneous cosmologies
there exists a preferred foliation of the 4-dimensional spacetime into
a product spacetime with a timelike Killing vector and a 3-dimensional
group of isometries $G_{3}$ acting simply transitively on spacelike
hypersurfaces such that the timelike Killing vector commutes with the
spacelike Killing vectors. Therefore we change from spacetime indices
to spatial indices.  We also adopt the comoving gauge whereby the
timelike basis vector is taken as parallel to the unit normal to the
surfaces of homogeneity. In general, the structure constants of the
symmetry groups of these different surfaces of constant time are
time-dependent
\cite{Ellis2006BianchiModelsThenAndNow,Taub1951,Heckmann1962Gravitation}.
However, from (\ref{tetradKV}), it follows that between homogeneous
slices, the structure constants are preserved up to a time-dependent
linear transformation. One is free to shift the time dependence
between the structure constants and the metric components by adjusting
the time evolution of the tetrad. We choose to put all the time
variation in the spatial metric components such that
\begin{equation}
   ds^{2}=dt^{2}-\gamma_{kl}(t)(e_{i}^{k}(x)dx^{i})(e_{j}^{l}(x)dx^{j}),
\label{metric}
\end{equation}
where $e_{i}^{k}(x)$ are the one-forms inverse to the spatial tetrad
which have the same structure constants $C_{\,\, ij}^{k}$ as the
isometry group and commute with the timelike unit normal
$\mathbf{e}_{0}$ to the surfaces of homogeneity:
$\mathbf{e}_{0}=\partial_{t}$ and
$\mathbf{e}_{i}=e_{i}^{j}\partial_{j}$, such that
$\left[\mathbf{e}_{i},\mathbf{e}_{j}\right]=C_{\,\,
  ij}^{k}\mathbf{e}_{k}$ and
$\left[\mathbf{e}_{0},\mathbf{e}_{i}\right]=0$. This is the approach
we will adopt for constructing the metric later, as the Einstein field
equations become ordinary differential equations in the metric
components $\gamma_{ij}(t)$. (Other approaches are based on the
automorphism group of the symmetry group
\cite{WainwrightEllis1997DynamicalSystemsInCosmology,
  CollinsHawking1973UniverseIsotropic} or put the time-dependence in
the commutation functions of the basis vectors \cite{Ellis1968}.)

We will now briefly review the Bianchi classification of $G_{3}$ group
types. The spatial part of the structure constants $C_{\,\, ij}^{k}$
can be decomposed into irreducible parts as follows
\begin{equation}
C_{\,\, ij}^{k}=\epsilon_{ijl}n^{lk}+a_{i}\delta_j^k-a_{j}\delta_i^k\label{decomp}\end{equation}
for $n^{ij}$ symmetric.  The Jacobi identities can then be rewritten
as
\begin{equation}
n^{ij}a_{j}=0.
\label{jacobi}
\end{equation}
Without loss of generality we can choose the tetrad so as to
diagonalise $n_{ij}=\diag \left(n_{1},n_{2},n_{3}\right)$ and to set
$a^{i}=\left(a,0,0\right)$ which reduces the Jacobi identities to
$n_{1}a=0$. This then allows one to classify the possible Lie
groups. We can define two broad classes of structure constants
according to whether $a=0$ (Class A) or not (Class B). See Table
\ref{tab: Bianchi} for more details concerning the individual Bianchi
types specified by the different options for the $n^{ij}$ matrix
entries.

If one were to put the time dependence in the structure constants
instead, one would of course have to show that the classification type
must be preserved by the evolution equations for $n(t)$ and $a(t)$. It
turns out that it is a generic property of the Einstein field
equations that they preserve symmetries in initial data within its
Cauchy development, so that the classification is in fact independent
of which approach one chooses \cite{HawkingEllis1973LargeScaleStructureOfSpaceTime}.

\begin{table}
\begin{centering}\begin{tabular}{|c|c|c|c|c|c|c|}
\hline
Class&
Type&
$n_{1}$&
$n_{2}$&
$n_{3}$&
\multicolumn{1}{c|}{$a$}&
\tabularnewline
\hline
\hline
$A$&
$\mathrm{I}$&
$\mathrm{0}$&
$\mathrm{0}$&
$\mathrm{0}$&
$\mathrm{0}$&
\tabularnewline
\hline
&
$\mathrm{II}$&
$\mathrm{+}$&
$\mathrm{0}$&
$\mathrm{0}$&
$\mathrm{0}$&
\tabularnewline
\hline
&
$\mathrm{VI_{0}}$&
$\mathrm{0}$&
$\mathrm{+}$&
$\mathrm{-}$&
$\mathrm{0}$&
\tabularnewline
\hline
&
$\mathrm{VII_{0}}$&
$\mathrm{0}$&
$\mathrm{+}$&
$\mathrm{+}$&
$\mathrm{0}$&
\tabularnewline
\hline
&
$\mathrm{VIII}$&
$\mathrm{-}$&
$\mathrm{+}$&
$\mathrm{+}$&
$\mathrm{0}$&
\tabularnewline
\hline
&
$\mathrm{IX}$&
$\mathrm{+}$&
$\mathrm{+}$&
$\mathrm{+}$&
$\mathrm{0}$&
\tabularnewline
\hline
\hline
$B$&
$\mathrm{V}$&
$\mathrm{0}$&
$\mathrm{0}$&
$\mathrm{0}$&
$\mathrm{+}$&
\tabularnewline
\hline
&
$\mathrm{IV}$&
$\mathrm{0}$&
$\mathrm{0}$&
$\mathrm{+}$&
$\mathrm{+}$&
\tabularnewline
\hline
&
$\mathrm{VI_{h}}$&
$\mathrm{0}$&
$\mathrm{+}$&
$\mathrm{-}$&
$\mathrm{+}$&
$h\equiv a^{2}/n_{2}n_{3}<0$\tabularnewline
\hline
&
$\mathrm{III}$&
$\mathrm{0}$&
$\mathrm{+}$&
$\mathrm{-}$&
$n_{2}n_{3}$&
\tabularnewline
\hline
&
$\mathrm{VII_{h}}$&
$\mathrm{0}$&
$\mathrm{+}$&
$\mathrm{+}$&
$\mathrm{+}$&
$h\equiv a^{2}/n_{2}n_{3}>0$\tabularnewline
\hline
\end{tabular}\par\end{centering}

\caption{\label{tab: Bianchi} Bianchi model classification.
Those containing FRW models as special cases are:
Bianchi $\mathrm{IX}$ (closed);
Bianchi $\mathrm{I}$ and Bianchi $\mathrm{VII_{0}}$ (flat);
Bianchi $\mathrm{V}$ and Bianchi $\mathrm{VII_{h}}$ (open).}

\end{table}

\section{Triaxial Bianchi IX Model\label{tri}}

We now consider Einstein-Hilbert gravity in a generic Bianchi IX model
with a minimally coupled scalar field. For generality we also include
a cosmological constant term. We therefore start with the action
\begin{equation}
S=\int{d^{4}x\sqrt{-g}\left[ \frac{1}{2\kappa} \left( R+2\Lambda
    \right)-\frac{1}{2}\nabla_{\mu}\phi\nabla^{\mu}\phi+V(\phi)
    \right]},
\label{action}
\end{equation}
where $R$ is the Ricci scalar, $\phi$ the scalar field and $V(\phi)$
its potential, which we will for simplicity assume to be that of a
simple massive scalar field, $V(\phi)=\frac{1}{2}m^2\phi^2.$ Variation
of the action with respect to the scalar field $\phi$ yields
conservation of energy-momentum
\begin{equation}
\nabla_{\mu}T^{\mu\nu}=0,
\label{genEMTcons}
\end{equation}
for the usual scalar field energy-momentum tensor
\begin{equation}
T_{\mu\nu}=\phi_{;\mu}\phi_{;\nu}
-g_{\mu\nu}\left({\textstyle\frac{1}{2}}\phi^{;\rho}\phi_{;\rho}-V(\phi)\right).
\label{scalarEMTV}
\end{equation}
Variation with respect to the metric yields the Einstein field
equations
\begin{equation} G_{\mu\nu}=\kappa T_{\mu\nu}+\Lambda
  g_{\mu\nu}.
\label{genEFE}
\end{equation}
We are working in Planck
units where $c=\hbar=1$ and we will set $\kappa=8\pi G$ to unity
eventually.

We choose to express the Bianchi IX metric in the form
\begin{equation}
ds^2=dt^2-{\textstyle\frac{1}{4}}\gamma_{i j }(t) \omega^i\omega^j
\label{lineelmt}
\end{equation}
for $\gamma_{i j }(t) =\diag \left(R_1^2(t), R_2^2(t), R_3^2(t) \right)$
and invariant 1-forms $\omega^i$ given by
\begin{eqnarray}
\omega^{1} & \equiv & dx+\sin y \,dz,\nonumber \\
\omega^{2} & \equiv & \cos x\,dy-\sin x\cos y\,dz,\nonumber \\
\omega^{3} & \equiv & \sin x\,dy+\cos x\cos y\,dz,
\label{inv1frms}
\end{eqnarray}
with corresponding Killing vectors
{(these just correspond to the rotations of the 3-sphere, i.e.
the Clifford translations)}
\begin{eqnarray}
\xi_{1} & \equiv & \sec y\cos z\,\partial_{x}+\sin z\,\partial_{y}-\tan
y\cos z
\,\partial_{z},\nonumber \\
\xi_{2} & \equiv & -\sec y\sin z\,\partial_{x}+\cos z\,\partial_{y}+\tan y\sin
z\,\partial_{z},\nonumber \\
\xi_{3} & \equiv & \partial_{z},
\label{kv}
\end{eqnarray}
and invariant basis
\begin{eqnarray}
X_{1} & \equiv & \partial_{x},\nonumber \\
X_{2} & \equiv & \sin x\tan y\,\partial_{x}
+\cos x\,\partial_{y}-\sin x\sec y\,\partial_{z},\nonumber \\
X_{3}& \equiv& -\cos x\tan y\,\partial_{x}+\sin x\,\partial_{y}+\cos x\sec
y\,\partial_{z}
\label{basis}
\end{eqnarray}
taken from \cite{Stephani2003ExactSolutions}. (We will consider an
alternative set of Killing vectors derived within the Conformal
Geometric Algebra framework
\cite{LasenbyLasenbyDoran2002ConformalGeometry,LasenbyDoran2003GeometricAlgebra}
related more obviously to an $(x,y,z)$ coordinate system in a future
work.)

Using the above expressions for the generating one-forms and expanding
out we get the following non-zero metric components for a triaxial
Bianchi IX universe (where  we have dropped the
explicit time dependence of the scale factors $R_i(t)$ for the sake
of brevity):
\begin{eqnarray}
g_{tt} & = & 1,\nonumber \\
g_{xx} & = & -{\textstyle\frac{1}{4}}R_{1}^{2}, \nonumber \\
g_{yy} & = & -{\textstyle\frac{1}{4}}(R_{2}^{2}\cos^{2}x+R_{3}^{2}\sin^{2}x),
\nonumber \\
g_{zz} & = & -{\textstyle\frac{1}{4}}
[R_{1}^{2}\sin^{2}y+(R_{2}^{2}\sin^{2}x+R_{3}^{2}\cos^{2}x)
\cos^{2}y],\nonumber \\
g_{xz}& = & -{\textstyle\frac{1}{4}}R_{1}^{2}\sin y,\nonumber \\
g_{yz}& = & -{\textstyle\frac{1}{4}}(R_{3}^{2}-R_{2}^{2})\sin x\cos x\cos y.
\label{trixmetric}
\end{eqnarray}

Introducing the usual definitions for the Hubble parameters
\begin{equation}
H_{i}(t)\equiv\frac{\dot{R_{i}}}{R_{i}}
\label{Hubble}
\end{equation}
for the three different directions, we find that for the above metric
the conservation of energy-momentum (\ref{genEMTcons})
corresponds to the Klein-Gordon-type equation
\begin{equation}
m^{2}\phi+\left(H_{1}+H_{2}+H_{3}\right)\dot{\phi}+\ddot{\phi}=0.
\label{generalscalarEoM}
\end{equation}
The $ti$-components of the Einstein field equations (for spatial index
$i$) give three dynamical equations for the Hubble parameters
$H_i$, the first of which reads
\begin{align}
2\dot{H}_1  = H_2H_3-2H_1^2-H_1H_3-H_1H_2+\Lambda-\kappa p \notag \\
            -\frac{5R_1^2}{R_2^2R_3^2} +\frac{3R_2^2}{R_3^2R_1^2}+\frac{3R_3^2}{R_1^2R_2^2}-\frac{6}{R_1^2} +\frac{2}{R_2^2}+\frac{2}{R_3^2},
\label{H1}
\end{align}
and the other two equations are obtained simply by swapping indices
$1\leftrightarrow 2$ and $1\leftrightarrow 3$.  The $tt$-component of
the Einstein field equations yields one Friedmann-type constraint
equation
\begin{align}
& -H_1H_2-H_2H_3-H_1H_3+\Lambda+\kappa\rho \notag\\
& +\frac{R_1^2}{R_3^2 R_2^2}+\frac{R_2^2}{R_3^2 R_1^2}+\frac{R_3^2}{R_1^2 R_2^2}
-\frac{2}{R_1^2}-\frac {2}{R_2^2}-\frac {2}{R_3^2} = 0,
\label{Fr}
\end{align}
where $\rho=\frac{1}{2}\dot{\phi^{2}}+V\left(\phi\right)$ and
$p=\frac{1}{2}\dot{\phi}^{2}-V\left(\phi\right)$ are the energy
density and pressure of the scalar field matter respectively.  Note
that there are no spatial gradients of $\phi$ in the expressions for
$\rho$ and $p$, by spatial homogeneity. {The case of vanishing potential
corresponds to a stiff fluid, $p=\rho$,  which has been investigated in
\cite{Barrow1978NatureQuiescentCosmology, Barrow1988prematureRecollapse}, amongst others. }

{For different applications, it might be useful to recast these
equations in terms of the averaged scale factor, i.e. the volume expansion,
\begin{equation}
R\equiv \left(R_1R_2R_3\right)^{\frac{1}{3}},
\label{volumeexp}
\end{equation}
its associated Hubble factor,
\begin{equation}
3H\equiv H_1+H_2+H_3,
\label{volumeexpH}
\end{equation}
and the shear $\sigma_{ij}$.
For example, this would be useful for separating the contributions from the
curvature and the shear. We display this form of the Einstein field
equations in the appendix. In the case of the biaxial Bianchi IX
model that we will  consider shortly, however, we believe
the parametrisation in terms of the different radii is  clearer.
When considering the cosmology at late times, however, we will find that our model
has isotropised sufficiently such that we can describe our model by an effective FRW-model
 with a scale factor given by the volume expansion of the Bianchi model.
}

It has been known for a long time that the Bianchi IX model (or indeed
any Bianchi model for which $C_{\,\, ij}^{j}=0$) is geodesically
incomplete and exhibits a curvature singularity for perfect fluid
matter (see,
e.g. \cite{RyanShepley1975HomogeneousRelativisticCosmologies}). The
singularity occurs precisely when the volume of an invariant
hypersurface goes to zero. The perfect fluid energy density can be
shown to diverge as $\rho_{fl}\sim\frac{1}{t}$ and the curvature
invariants are singular as well, as $t\rightarrow 0$.  However, it is
the oscillatory character of the singularity that makes perfect fluid
matter in a Bianchi IX universe model unsatisfactory as a cosmological
model: the Big Bang is an essential singularity.  Oscillations in the
ratios of the different scale factors in $\ln (t)$ effectively show
that there is an infinite history upon approaching the Big Bang,
making it impossible to trace back to it.  Conversely, coming out of
an essential singularity to reach the observed universe is ill-defined
for the same reasons. This also holds, in particular, for the biaxial
case by virtue of the general theorem. We will see below that for a {biaxial}
Bianchi IX model with scalar field matter, however, the picture will
be qualitatively different.

Scalar field matter is {still} unsatisfactory in the full triaxial Bianchi IX
model, {however, } since it exhibits the same behaviour outlined above.
{As we show below, it is only when we impose axial symmetry, that a solution is}
possible in which all physical quantities such as energy density and
curvature invariants remain finite at the Big Bang, and the universe
extends smoothly across what is no longer an essential singularity
into a well-behaved pre-Big Bang phase (though parity-inverted!). This
then raises the question as to whether the axially symmetric case is
stable to small pertubations in biaxiality, which we will address in
Section \ref{pert}.

\section{Biaxial Bianchi IX Model\label{axi}}

We therefore now specialise to the case where two of the axes are
equal, $R_2(t)=R_3(t)$, which leads to the simplified metric (again
with the $t$-dependence of the scale factors suppressed)
\begin{equation}
g_{\mu\nu}\equiv\left(\begin{array}{cccc}
1 & 0 & 0 & 0\\
0 & -\frac{1}{4}R_{1}^{2} & 0 & -\frac{1}{4}R_{1}^{2}\sin y\\
0 & 0 & -\frac{1}{4}R_{2}^{2} & 0\\
0 & -\frac{1}{4}R_{1}^{2}\sin y & 0 &
-\frac{1}{4}(R_{1}^{2}\sin^{2}y+R_{2}^{2}\cos^{2}y)\end{array}\right).
\label{specialmetric}
\end{equation}

Now there are only two dynamical equations, for the two non-degenerate
Hubble parameters, and one Friedmann constraint.  The first dynamical
equation is
\begin{equation} 2\dot{H}_{2}+3H_{2}^{2}+\kappa
  p-\Lambda=\frac{1}{R_{2}^{2}}\left(3\frac{R_{1}^{2}}{R_{2}^{2}}-4\right)
\label{special1}
\end{equation}
and the other reduces to
\begin{equation}
  2\dot{H}_{1}+2H_{1}^{2}-H_{2}^{2}+2H_{1}H_{2}+\kappa
  p-\Lambda=-\frac{1}{R_{2}^{2}}\left(5\frac{R_{1}^{2}}{R_{2}^{2}}-4\right).
\label{special2}
\end{equation}
In the isotropic limit, these equations reduce to appropriate
combinations of the usual Friedmann and acceleration equations, as
required.  The Einstein field equations further yield the Friedmann
constraint
\begin{equation}
  H_{2}^{2}+2H_{1}H_{2}-\kappa\rho-\Lambda=\frac{1}{R_{2}^{2}}
  \left(\frac{R_{1}^{2}}{R_{2}^{2}}-4\right).\label{specialFr}
\end{equation}
This is straightforwardly seen to reduce to the standard Friedmann
equation as $R_1\rightarrow R_2=R_3$. The equations of motion for the
matter content follow straightforwardly from energy-momentum
conservation as above. In fact they are easily deduced from the
triaxial case and read
\begin{equation}
  m^{2}\phi+\left(H_{1}+2H_{2}\right)\dot{\phi}+\ddot{\phi}=0.
\label{specialEoM}
\end{equation}
We prefer the viewpoint from which the equations
(\ref{special1})-(\ref{special2}) are dynamical equations for the
Hubble parameters and regard the scale factors $R_{i}$ as derived
quantities, but these equations are obviously equivalent to second
order equations in terms of the radii.

The simplicity of these equations suggests that a relatively
simple solution should be possible. {We are particularly interested {here} in
solutions with definite parity,
{and defer discussion of solutions with indefinite parity to a future work}.
We show below that both odd-parity and even-parity
series expansions exist around the `Big Bang', which are therefore valid
starting points for numerical integration}, but first we
briefly discuss the generation of a family of solutions from a given solution
by scaling.

\subsection{Families of solutions related by scaling}

Given a solution to the equations (\ref{special1})-(\ref{specialEoM}),
a family of solutions is generated by scaling with a constant $\sigma$
and defining
\begin{equation}
\bar{R_i}(t)=\frac{1}{\sigma} R_i(\sigma t),\,\,\, \bar{H_i}(t)=\sigma H_i(\sigma t),\,\,\, \bar{\phi}(t)=\phi(\sigma t)
,\,\,\, \bar{m}=\sigma m , \,\,\, \bar{\Lambda}=\sigma ^2\Lambda.
\label{scaling}
\end{equation}
(This scaling is in fact analogous to the one found previously in
\cite{LasenbyDoran2005ClosedUniversesdSandInflation}.) This scaling
property is valuable for numerical work, as a range of situations can
be covered by a single numerical integration. Furthermore, many
physically interesting quantities turn out to be invariant under
changes in scale. This scaling property does not, however, survive
quantisation, so one would have to be careful when considering vacuum
fluctuations.

\subsection{{Odd-parity} series expansion around the {`Big Bang'}}\label{series}

It is well known \cite{Ellis2006BianchiModelsThenAndNow,
  RyanShepley1975HomogeneousRelativisticCosmologies} that pancake
singularities (which we will identify with the {`Big Bang'}, and choose to
happen at time $t=0$), where one radius goes to zero and the other two
remain finite as $t\rightarrow 0$, can occur in Bianchi models.  This
is already in some sense an improvement over the FRW case, where the
singularity is pointlike. However, pancake singularities are known to
occur in Bianchi I (as well as cigar singularities), whereas Bianchi
IX is commonly thought generically to exhibit oscillatory
singularities. We will now show that, with scalar field matter, the
pancake singularity in the Bianchi IX model is not in fact a curvature
singularity at all, because all physical quantities remain finite as
$t\rightarrow 0$ and extend smoothly into a parity-inverted universe
for $t<0$.

We assume that it is the non-degenerate radius $R_{1}$ that tends to
zero as $t\rightarrow 0$. Close to the Big Bang, the linear term
becomes dominant for $R_1(t)$, whereas the other radii
$R_{2}(t)=R_{3}(t)$ tend to a non-zero constant, as does the scalar
field $\phi(t)$. Indeed, for a series solution ansatz in which we
assume oddness for $R_{1}$ and evenness for the other functions, i.e.
\begin{align}
R_{1}(t)&=t\left(a_{0}+a_{2}t^{2}+a_{4}t^{4}+\dots\right)\notag\\
R_{2}(t)=R_{3}(t)&=b_{0}+b_{2}t^{2}+b_{4}t^{4}+\dots \,\,\,\,, \notag\\
\phi(t)&=f_{0}+f_{2}t^{2}+f_{4}t^{4}+\dots\notag\\
\label{seriesansatz}
\end{align}
then the three dynamical equations (\ref{special1}), (\ref{special2})
and (\ref{specialEoM}) allow us to fix the three series term-by-term,
given the initial values $a_{0}=\dot{R}_{1}(0)$,
$b_{0} = R_{2}(0)$ and $f_{0} = \phi(0)$. The fact that this also
satisfies the Friedmann energy constraint (\ref{specialFr}) then
proves that this {odd-parity series solution} is a valid expansion around the Big Bang, which we
can use as a starting point for numerical integration.

Intriguingly, it turns out that the spacetime is non-singular insofar
as the Riemann tensor is well-behaved at the Big Bang and  so
are all curvature invariants. In fact only three components of the
Riemann tensor are non-zero at $t=0$:
\begin{equation}
R_{tyty}={\textstyle\frac{1}{16}}\kappa m^{2}f_{0}^{2}b_{0}^{2}
+{\textstyle\frac{1}{8}}\Lambda
  b_{0}^{2}-{\textstyle\frac{1}{2}},\,\, \,  R_{tztz}=R_{tyty}
\cos^{2}y, \,\,
\, R_{yzyz}=-{\textstyle\frac{1}{4}}b_0^2\cos^{2}y.
\label{Riemann}
\end{equation}

The fact that there is no curvature singularity means that these
series solutions can be continued through to negative values of
$t$. That is, $R_2=R_3$ and $\phi$ are even for $t<0$, whereas $R_1$
is odd i.e. negative for $t<0$. This amounts to a parity inversion at
the Big Bang. Instantaneously, as the hypersurface volume goes to zero
at the Big Bang, the spatial hypersurface is 2-dimensional.  The fact
that the universe can be momentarily `dimensionally reduced' is
interesting. Nonetheless, even though the hypersurface volume goes to
zero instantaneously, information about the evolution is encoded in
the derivatives.  If one adopts the string-inspired viewpoint
\cite{BBS, Ortin, Dbranes} that the universe could be described by a
3-brane, the instantaneous conversion of 3-branes into 2-branes seems
problematic in type IIA and IIB String theory, but could be
interesting from a type IIA/IIB String theory duality or AdS/CFT point
of view.  From a more conservative perspective, this process resembles
Taub-NUT space
\cite{RyanShepley1975HomogeneousRelativisticCosmologies, Ortin}, which
is a biaxial Bianchi IX vacuum solution that can be represented as a
disc {that} evolves into an ellipsoid and back into a disc. It
therefore also shows the same feature of dimensional reduction as our
model, and moreover also does not have a geometric singularity during
this collapse. In particular, it is thought to evolve from timelike
open sections in a NUT region, via lightlike sections (called Misner
bridges), to spacelike closed sections in the Taub region, back into
timelike open sections in the other NUT region. This
open-to-closed-to-open transition is not mathematically singular, but
it is incomplete, as geodesics spiral infinitely many times around the
topologically closed spatial dimension as they approach the boundary
\cite{KonkowskiHelliwellShepley1985quasiregularI,
  KonkowskiHelliwell1985quasiregularII}. This type of singularity is
called `quasiregular' in the Ellis and Schmidt classification
\cite{EllisSchmidt1977SingularSpacetimes}
{(these include the well-known `conical' singularities \cite{KonkowskiHelliwell2005Classification})},
 as opposed to a (scalar or non-scalar) curvature singularity. {For completeness,
 a possible parametrisation for the Taub-NUT metric is
\begin{equation}
ds^2=2dt\omega^1-{\textstyle\frac{1}{4}} R_1^2 (\omega^1)^2-{\textstyle\frac{1}{4}}R_2^2\left[(\omega^2)^2+(\omega^3)^2\right]
\label{lineelmttaubnut}
\end{equation}
as compared with
\begin{equation}
ds^2=dt^2-{\textstyle\frac{1}{4}} R_1^2 (\omega^1)^2-{\textstyle\frac{1}{4}}R_2^2\left[(\omega^2)^2+(\omega^3)^2\right]
\label{lineelmtus}
\end{equation}
in our model. The interested reader may refer to \cite{RyanShepley1975HomogeneousRelativisticCosmologies},
for example, for a discussion of Taub-NUT.
}

It is worth noting that in deriving the above series solution, one can
instead start with a general Taylor series expansion in the full
triaxial case. Expanding around the Big Bang rather than any other
point amounts to demanding that the constant term in the Taylor
expansion for $R_1$ be zero (in fact, this could be any radius, as
$R_1$ is not special in the triaxial case), whereas everything else is
in principle undetermined.  Imposing the dynamical equations, however,
forces the two non-zero radii to be the same term-by-term for the
expansion to be valid, i.e. a Big Bang-like expansion only works for
the axisymmetric model; there is an essential singularity unless we
consider the special, biaxial case. The validity of the equations of
motion further implies the above mentioned even form for $R_2=R_3$ and
$\phi$, and odd behaviour for $R_1$.

Since the existence of the above series expansion around the Big Bang
demonstrates that an axisymmetric pancake singularity is a valid
starting point for numerical integration, we therefore solved the
biaxial equations numerically subject to the appropriate boundary
conditions. We found good agreement with the series solutions within
the range of their validity.  The large parameter space of this model
admits both viable {(in the sense of compatible with present observations)}
and unrealistic {(in the sense of incompatible with present observations)} cosmologies. We defer seeking
realistic model parameters and displaying the numerical results until
Section~\ref{cos} and instead concentrate next on addressing the issue
of geodesic completeness.

\subsection{Behaviour of geodesics\label{geodesics}}

As we have shown, our model has no curvature singularity at the Big
Bang, at which all physical quantities remain finite. Thus, at the
level of the evolution equations, the model is well-behaved.  We now
consider the question of geodesic completeness of our model to
determine whether it possesses a non-curvature singularity at $t=0$.

The geodesic equations are most easily obtained using the Lagrangian
formalism, in which
\begin{equation}\label{GXL}
L = {\textstyle\frac{1}{4}}g_{\mu\nu}\dot{x}^\mu \dot{x}^\nu
\end{equation}
is varied with respect to the coordinates $[x^\mu]=(t, x, y, z)$; here
a dot denotes a derivative with respect to some affine parameter
$\lambda$ and the factor $\frac{1}{4}$ is included for later
convenience. Inserting the biaxial metric (\ref{specialmetric}) into
$L$ yields
 \begin{equation}
\label{GXLex}
L=4\dot{t}^2-R_1^2\dot{x}^2-2R_1^2\sin y \dot{x}\dot{z}-R_2^2\dot{y}^2
-\left(R_1^2\sin^2y+R_2^2\cos^2y
\right)\dot{z}^2.
\end{equation}
Since $L$ is independent of the $x$- and $z$-coordinates, the
corresponding Euler--Lagrange equations yield two conserved quantities
$K_x$ and $K_z$ according to the relations
\begin{eqnarray}
-2R_1^2\left(\dot{x}+\sin y \dot{z}\right) & \equiv & K_x, \label{Kx} \\
-2\left[R_1^2 \sin y\left(\dot{x}+\sin y \dot{z}\right)+R_2^2 \cos^2y
-\dot{z}\right] & \equiv & K_z, \label{Kz}
\end{eqnarray}
which may be solved for $\dot{x}$ and $\dot{z}$ to yield
\begin{eqnarray}
    \dot{x} & = & \frac{K_zR_1^2 \sin y -K_x\left(R_1^2\sin^2y+R_2^2\cos^2y
    \right)}{2R_1^2 R_2^2 \cos^2 y}, \label{xdot}\\
\dot{z}&=&\frac{K_x\sin y  - K_z}{2R_2^2 \cos^2 y}. \label{zdot}
\end{eqnarray}
Substituting these expressions back into the Lagrangian,
the Euler--Lagrange equation for $y$ then reads
\begin{equation}\label{ELy}
    4R_2^3 \cos^3y\left(R_2\ddot{y}+2\frac{\partial R_2}{\partial t}\dot{t}\dot{y}\right)=
    \left(K_x^2+K_z^2\right)\sin y -2K_xK_z+K_xK_z \cos^2 y.
\end{equation}
%


Let us first consider null geodesics for which
$g_{\mu\nu}\dot{x}^\mu\dot{x}^\nu=0=L/4$.  To begin with, we wish only
to show that some geodesics extend smoothly through the Big Bang, and
therefore we can choose a convenient form. Inspection of equations
(\ref{xdot}) and (\ref{zdot}) shows that we cannot set either of
$\dot{x}$ and $\dot{z}$ alone to zero by choosing suitable constants,
but we can consider the case where both vanish (this requires the
choice $K_x=K_z=0$ and essentially corresponds to motion in the
$y$-direction).  In this case the Euler--Lagrange equation for $y$
simplifies to $d\left(R_2^2\dot{y}\right)/d\lambda=0$ which can be
immediately integrated to yield
\begin{equation}\label{ELy2}
R_2^2\dot{y} =  K_y,
\end{equation}
where $K_y$ is another constant of the motion. The constraint that the
geodesic is null now reads $4\dot{t}^2-R_2^2\dot{y}^2=0$, which gives
\begin{equation}\label{dtdy}
    \frac{dy}{dt}=\frac{\dot{y}}{\dot{t}}=\pm\frac{2}{R_2}.
\end{equation}
This is reminiscent of the Friedmann case, but one must now bear in
mind that in our model $R_2$ is finite and non-zero at the Big Bang in
contrast to the scale factor in an FRW model. Thus, there is no longer
a difficulty in integrating through the Big Bang; rather, it is
trivial. Indeed, looking at the Euler--Lagrange equation for $y$
directly in this case, one has
\begin{equation}\label{ELy3}
R_2\ddot{y}+2\frac{\partial R_2}{\partial t}\dot{t}\dot{y}=0,
\end{equation}
which amounts to $\ddot{y}=0$ near the Big Bang where $R_2$ has a
minimum, also suggesting that such a light ray goes straight through
the Big Bang. Thus photon motion in $y$ is effectively undisturbed by
the pancaking.

Let us now consider a photon moving in the $x$-direction.  It is
easiest to see from (\ref{ELy}) that if a photon starts at $y=0$ it
will stay there. Hence for $K_z\equiv 0$ and $y=0$, the results
(\ref{xdot}), (\ref{zdot}) and (\ref{ELy}) yield $\dot{z}=0$,
$\dot{y}=0$ and
\begin{equation}\label{xpxdot}
    \dot{x}=-\frac{K_x}{2R_1^2}.
\end{equation}
In this case, the constraint that the geodesic is null yields
$4\dot{t}^2=R_1^2\dot{x}^2$. Hence, close to $t=0$, where $R_1$ is
approximately linear in $t$, once has
\begin{equation}\label{xpxdotovertdot}
    \frac{dx}{dt}=\frac{\dot{x}}{\dot{t}}=\pm\frac{2}{R_1}\propto\pm
    \frac{1}{t}.
\end{equation}
Integrating we obtain
\begin{equation}\label{xpxoft}
    x(t)=c_1\ln (-t)+c_2 \mathrm{\,\,\,\,\,\, or \,\,\,\,\,\,}
x(t)=c_3\ln (t)+c_4,
\end{equation}
for some constants $c_i,\,\,\, i=1\dots 4$, which shows that at $t<0$
we must have the former solution and at $t>0$ the latter.

Let us illustrate these solutions by considering the specific case of
a photon travelling in the positive $x$-direction as time increases,
i.e. $\frac{dx}{dt}>0$.  For $t<0$ we have
$\frac{dx}{dt}=\frac{c_1}{t}$ and for this to be positive for negative
$t$ we must have $c_1<0$.  This also yields $R_1\approx
-\frac{2}{c_1}t$ from (\ref{xpxdotovertdot}) which is negative for
$t<0$ as expected.  For $t>0$ (\ref{xpxoft}) yields
$\frac{dx}{dt}=\frac{c_3}{t}>0$ so $c_3>0$ and $R_1\approx
\frac{2}{c_3}t$. So it seems that the photon $x$-coordinate goes to
$+\infty$ as $t \to 0$ from below, reappearing just after $t=0$ at
$x=-\infty$.  This is less problematic than it seems: the spatial
sections in Bianchi IX are topologically $S^3$, so it is natural to
think of the coordinates as angles that should be periodically
identified. Furthermore, at $t=0$ and therefore $R_1=0$,
(\ref{metric}) yields vanishing proper distance between different
values of $x$, which presumably means that this direction has shrunk
to a point. Since our coordinates are angles, the result of the photon
going to $\infty$ appears to signify that the geodesics are infinitely
spiralling (which is reminiscent of Taub-NUT) around a closed spatial
dimension as it is collapsing.  The proper distance traversed and time
taken both, however, become infinitesimally small, so the whole process
might still be finite.  Thus it is not clear that anything singular
has in fact happened to the photon trajectory, despite appearances to
the contrary; this will also be discussed further in a future work.

We will now consider massive particles, which travel along timelike
geodesics. The only difference in this case is that one must now impose
the timelike geodesic normalisation constraint
$g_{\mu\nu}\dot{x}^\mu\dot{x}^\nu=1=L/4$. Let us again begin by
considering a particle travelling in the $y$-direction. In this case,
(\ref{ELy2}) still holds, but the normalisation constraint becomes
$4\dot{t}^2=4+R_1^2\dot{y}^2$. This yields
\begin{equation}\label{ymdybydt}
    \frac{dy}{dt}=\pm\frac{2K_y}{\sqrt{4R_2^2+K_y^2}}.
\end{equation}
At $t=0$, the right-hand side is still constant to first-order,
mirroring the massless case, so massive particles motion in the
$y$-direction is also essentially unaffected by the pancaking.

Finally, we consider motion of a massive particle along the
$x$-direction. Again (\ref{xpxdot}) holds, but the normalisation
constraint is now $4\dot{t}^2=4+R_1^2\dot{x}^2$,
which yields
\begin{equation}\label{xmxdotovertdot}
\frac{dx}{dt}=\frac{\dot{x}}{\dot{t}}=\mp\frac{2K_x}{R_1\sqrt{16R_1^2+K_x^2}}.
\end{equation}
Since $R_1\approx Ct$ near $t=0$, for a constant $C$, on integrating
we obtain
\begin{equation}\label{xmxoft}
x(t)
=\frac{2}{C}\ln\left[\pm\frac{t}{2K_x\left(K_x+\sqrt{K_x^2+16C^2t^2}\right)}
\right],
\end{equation}
In this regime the denominator in the logarithm is approximately
constant so massive particles will display qualitatively precisely the
same behaviour as photons as they approach and leave the point of
pancaking.

Thus, in summary, we find that there are null and timelike geodesics
that go smoothly through the Big Bang into a pre-Big Bang phase, but
other such geodesics that spiral infinitely around a topologically
closed spatial dimension. This issue does not seem to warrant too much
concern for our model, owing to the periodic identification of these
angular coordinates on the 3-spheres. This further supports our claim
that this model is substantially better behaved than most conventional
cosmological models. The issue of geodesic completeness will be
further discussed in a future paper, but it is worth noting here that,
even if the spacetime is incomplete, the singularity at the Big Bang
can, at worst, be of quasiregular type
\cite{EllisSchmidt1977SingularSpacetimes} as occurs also in Taub-NUT
\cite{KonkowskiHelliwellShepley1985quasiregularI,
  KonkowskiHelliwell1985quasiregularII}.

\section{Realistic biaxial Bianchi IX cosmology}\label{cos}

We now examine the viability of our model for providing a realistic
cosmology. We begin by noting the argument of
\cite{LasenbyDoran2005ClosedUniversesdSandInflation,
  LasenbyDoran2004ConformalModelsICsCMB}, that there is a natural,
geometrical boundary condition on the universe resulting from the need
to match a Big Bang phase onto an asymptotic de Sitter phase within a
particular type of conformal representation.  Here a genuine
cosmological constant $\Lambda$ is being assumed, rather than a
quintessence model.  Using a conformal embedding and the symmetries of
de Sitter space, a boundary condition is arrived at that the total
elapsed conformal time should be equal to $\frac{\pi}{2}$. The
condition is simply to demand that the future asymptotic de Sitter
state be the future infinity surface of the conformal embedding. This
amounts to imposing a boundary condition at temporal infinity, much
like in (quantum) field theory.  This singles out a particular flow
line in $\left(\Omega_{m},\Omega_{\Lambda}\right)$ space.

In the context of a closed FRW model, it is shown in
\cite{LasenbyDoran2005ClosedUniversesdSandInflation,
  LasenbyDoran2004ConformalModelsICsCMB} that the conformal time
constraint predicts within the correct range the degree of flatness of
the universe, as well as the size of the cosmological
constant. Moreover, the computed inflationary perturbations were shown
to be consistent with WMAP data and could even account for the
low-$\mathit{\ell}$ deficit. Extending this conformal time constraint
to our Bianchi model, we find that this constraint can be satisfied by
setting the free parameters in our model to  $\kappa=1$,
$m=\frac{1}{64000}$, $a_{0}=1.2$, $b_{0}=18000$ and $f_{0}=13$. (These values
are for $\kappa$ set to 1, and are essentially a representative set, rather
than having been fixed to get best agreement with current data.)  These parameter
values are all `natural', but in order to fix the normalisation of the
perturbation spectrum, the mass of the scalar field had to be rescaled
and $b_0$ changes to a less natural value accordingly.  This choice
for the mass of the scalar field needs to be put in by hand for every
model so does not constitute any unusual fine-tuning.

This model looks like the universe that we observe in the sense that
it comes out of a Big Bang-like state,
followed by an inflationary phase and eventually reaches a state of
steady expansion (see Figure \ref{figpanc}).  There it
can feasibly be matched onto a model of radiation domination followed
by matter domination to recover our standard cosmology.  Note,
however, that this Bianchi model is interesting irrespective of our
arriving at the particular parameter values above by using the
conformal time condition.

\condcomment{\boolean{includefigs}}{
\condcomment{\boolean{tikzfigs}}{
\begin{figure}
\tikzstyle{background grid}=[draw, black!50,step=.5cm]
\begin{tikzpicture}
\node (img) [inner sep=0pt,above right]
{\includegraphics[width=8cm]{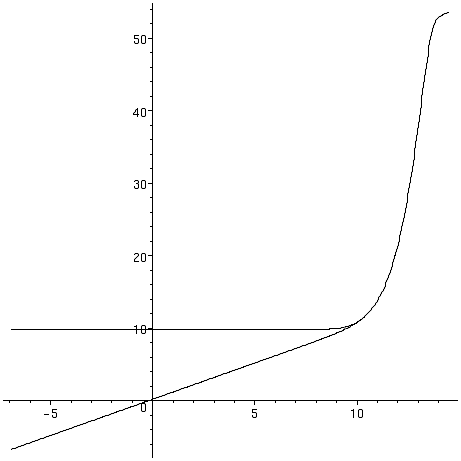}};
\draw (7:5.5cm) node {$\ln(t)$};
\draw (110:2.5cm) node {$\ln\left(R_2(t)\right)$};
\draw (162:0.7cm) node {$ \ln\left(R_1(t)\right)$};
\end{tikzpicture}
\caption[dummy1]{Dynamics of the biaxial Bianchi IX model:
 evolution of the logarithm of the scale factors $R_1$ and $R_2$ in Planck
lengths $l_p$ versus log time ($t$ in units of Planck time $t_p$).}
\label{figpanc}
\end{figure}}

\condcomment{\boolean{psfigs}}{
\begin{figure}
\includegraphics[width=8cm]{Bianchi_fig1_panc1}
\caption[dummy1]{Dynamics of the biaxial Bianchi IX model:
 evolution of the logarithm of the scale factors $R_1$ and $R_2$ in Planck
lengths $l_p$ versus log time ($t$ in units of Planck time $t_p$).}
\label{figpanc}
\end{figure}}
}

As mentioned earlier, the model exhibits no curvature singularity,
with all physical quantities remaining finite through the Big Bang.
The fact that the vanishing radius is odd in time results in parity
inversion as we go through the Big Bang. The property that the energy
density remains finite is contrary to the most common scenarios
(although it was previously known that a massive scalar field can lead
to a non-singular bounce in the special case of a closed FRW universe;
see \cite{GordonTurok2003PerturbationsThroughBounce} for more
information on the history of such models).

Figure \ref{figpanc} demonstrates that the scale factor $R_1(t)$
approaches zero linearly as $t \rightarrow 0$, as expected from the
series expansion in Section \ref{series}.  Note that this behaviour
results in a slope of 1 in the log-log plot as $\ln t\rightarrow
-\infty$.  The other radii $R_2=R_3$, as well as physical variables
such as the scalar field and the scalar field energy density, tend to
a constant at the Big Bang as shown in Figure \ref{figpanc}, confirming
our results from the series expansion. This shows that this model
exhibits a pancake singularity. The relaxation of the assumption of
isotropy makes a variety of singularities possible which generalise
the pointlike FRW-singularity.  For instance, pancake, barrel and
cigar singularities are known to occur in Bianchi models
\cite{Ellis2006BianchiModelsThenAndNow,
  RyanShepley1975HomogeneousRelativisticCosmologies}.  (In both barrel
and cigar singularities, two scale factors tend to zero but the third
increases without bound in cigar singularities, and approaches a
constant in barrel singularities.)  It is worth noting, however, that
a pancake singularity is atypical for Bianchi IX models.

\condcomment{\boolean{includefigs}}{
\condcomment{\boolean{tikzfigs}}{
\begin{figure}
\tikzstyle{background grid}=[draw, black!50,step=.5cm]
\begin{tikzpicture}
\node (img) [inner sep=0pt,above right]
{\includegraphics[width=8.0cm]{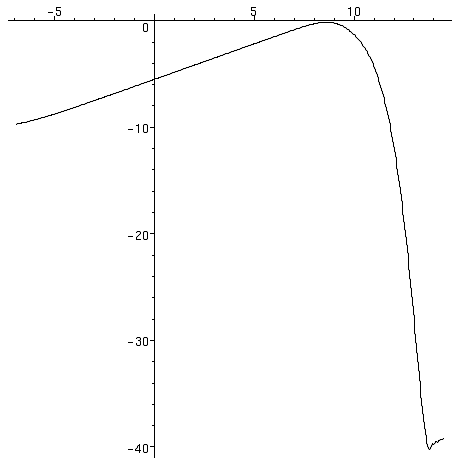}};
\draw (57:10cm) node {$\ln(t)$};
\draw (78:5cm) node {$-\ln\left(R(t)H(t)\right)$};
\end{tikzpicture}
\caption[dummy4]{Dynamics of the biaxial Bianchi IX model: evolution
  of the logarithm of the comoving Hubble radius versus log time. Note
  that the curvature radius $R(t)\equiv
  \left(R_1R_2R_3\right)^{\frac{1}{3}}$, with corresponding Hubble
  function $H(t)\equiv \frac{\dot{R}}{R}$ and associated Hubble radius
  $H^{-1}$. The parameters for the model are described in the
  text. During inflation, $1/(RH)$ decreases with time, corresponding
  to accelerated expansion. Thus this model leads to a period of
  inflation lasting approximately for the period $\ln t = 9 -
  13.5$. All quantities are in Planck units.  }
\label{fig4}
\end{figure}}

\condcomment{\boolean{psfigs}}{
\begin{figure}
\includegraphics[width=8.0cm]{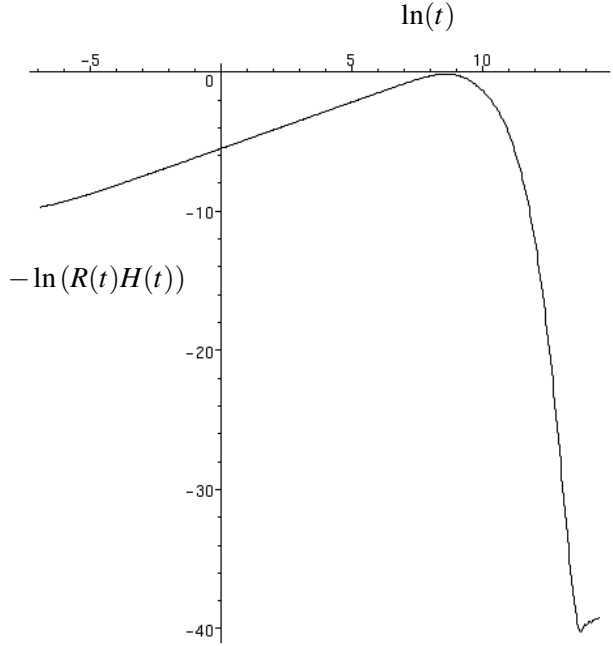}
\caption[dummy4]{Dynamics of the biaxial Bianchi IX model: evolution
  of the logarithm of the comoving Hubble radius versus log time. Note
  that the curvature radius $R(t)\equiv
  \left(R_1R_2R_3\right)^{\frac{1}{3}}$, with corresponding Hubble
  function $H(t)\equiv \frac{\dot{R}}{R}$ and associated Hubble radius
  $H^{-1}$. The parameters for the model are described in the
  text. During inflation, $1/(RH)$ decreases with time, corresponding
  to accelerated expansion. Thus this model leads to a period of
  inflation lasting approximately for the period $\ln t = 9 -
  13.5$. All quantities are in Planck units.  }
\label{fig4}
\end{figure}}

}

The evolution of the oblateness $R_2(t)/R_1(t)$ (see Figure
\ref{figpanc}), suggests that the evolution equations favour the radii
$R_2(t)=R_3(t)$ becoming similar to the radius $R_1(t)$, in agreement
with the need to be tolerably close to an FRW cosmology at late times.
(This ties in nicely with the fact demonstrated in Section \ref{pert}
that the biaxiality of our model, i.e. $R_2(t)=R_3(t)$, is also stable
to sufficiently small perturbations of the type
$R_3(t)/R_2(t)=1+\delta(t)$, thereby allowing an almost-FRW model in
which all axes are similar.) Note, however, that Figure \ref{figpanc}
demonstrates that the universe was significantly anisotropic until at
least $\ln t \approx 10$.  The scalar field seems to isotropise the
universe and subsequently drive inflation. (For related literature on
isotropisation, see \cite{Wald1983AsymptoticBehavior, Barrow1987Nohair,
Maeda1992PowerLaw, Maeda1993NoHairBianchi, Salucci1983GeneralBianchiAdiabatic}.)

We can define an averaged
scale factor by $R(t)\equiv \left(R_1R_2R_3\right)^{\frac{1}{3}}$,
with which we can associate an averaged Hubble function $H(t)\equiv
\dot{R}/R$, as usual, and thereby define a comoving Hubble radius
$1/(RH)$.  Figure \ref{fig4} shows the evolution of the comoving
Hubble radius $1/(RH)$ for our model.  Inflation is a period of
accelerated expansion, which results in the comoving Hubble radius
$1/(RH)$ decreasing with time. From the plot we can infer that for our
model a period of inflation does occur and lasts approximately from
$\ln t \approx 9$ to $\ln t \approx 13.5$. During this inflationary
period $\ln(R)$ increases from about 10 to 55 (c.f. Figure
 \ref{figpanc}), corresponding to roughly 45 e-folds of inflation.

In order to produce the fluctuation spectrum observed in our own
universe, it is thought that the present Hubble horizon must have been
inflated by at least about 40-50 e-folds, that is, 40-50 e-folds
between the time where the present Hubble scale exited the horizon and
the end of inflation
\cite{LiddleLyth2000CosmologicalInflationAndLargeScaleStructure,
  Ellis2003WMAPandCurvature}. This minimal number of e-folds would
lead to a universe that departs from spatial flatness by an amount
that is just visible today \cite{Starobinsky96spectrum}. If more time
was spent on the inflationary attractor, the universe would be driven
closer to flatness and might in fact become indistinguishably close to
spatially flat.

Since our model produces around 45 e-folds of inflation, for agreement
with observations, we would therefore require that the present horizon
scale leave the Hubble horizon soon after the onset of inflation. This
is in agreement with Uzan, Kirchner and Ellis' estimates
\cite{Ellis2003WMAPandCurvature} and, in particular, also leads to a
visibly closed geometry (cf. Starobinsky
\cite{Starobinsky96spectrum}).

Figures \ref{figpanc} and \ref{fig4} show that the phases of
isotropisation and inflation might have overlapped in the time period
$\ln t \approx 9-10$ such that structure on the largest scales could
have been laid down whilst the universe was still significantly
oblate. In particular, we have just argued that the present horizon
scale should have crossed the horizon around that time. So imprints of
oblateness might actually be experimentally accessible to us. Such
considerations, including perturbation analysis, CMB imprints,
etc. will be described in future work. Clearly, a proper analysis must
generalise standard inflationary cosmological computations
\cite{HobsonLasenbyEfstathiou2006GeneralRelativity,
  LiddleLyth2000CosmologicalInflationAndLargeScaleStructure} to
anisotropic cosmologies \cite{Lyth2008statisticalAnisotropy}.
{In particular, there are subtleties regarding the correct choice of vacuum
(generalisation of the Bunch-Davies vacuum), the definition of the canonical variables
(generalisation of the Mukhanov-Sasaki variables), the discreteness of the eigenmodes of the
Laplacian on the 3-sphere (as mentioned above, we use a continuum approximation)
as well as the definition of $k^2$, which arises since the direction  of the
wavevector $k_i$ starts to matter due to the anisotropy. Since inflation drives
the spacetime closer to flatness, it might also be possible to approximate the spacetime
by a Bianchi I model and  quantise the perturbations in this approximation as an intermediate step.
There has been recent progress in the cosmological perturbation theory of Bianchi I models,
to which the interested reader may wish to refer \cite{PitrouUzan2007CosmoPerturbationsAnisotropic,
Pitrou2008PredictionsAnisotropicEra, Contaldi2007InflationaryPertubationsInBianchi}.
Their results further show that the two gravitational wave polarisations do not necessarily
have the same power spectra. The spectra also do not seem to reduce to the standard results
in the limit of vanishing shear. At very
early times, when the shear dominates over the curvature, the Bianchi IX model is also very close to a Bianchi I model
 -- though not topologically, of course. Note also that since the issue of how geodesics
 propagate through the pancaking is as yet not completely resolved, we only consider perturbations
 `this side' of the pancaking, rather than trying to track or match perturbations from both sides
 \cite{GordonTurok2003PerturbationsThroughBounce,
 GrattonTurok2003ConditionsForGeneratingScaleInvariantDensityPerturbations,
 GrattonTurok2006CosmicPerturbationsCyclicAges}.}

As a first approximation, however, one may
consider the slight anisotropy during inflation as a perturbation to
standard {FRW} results and neglect it to zeroth order.
{In this case, the definition of a comoving wavevector $k$ becomes unambiguous.} A detailed
computation would {ultimately have to validate this approach}. In
this approximation the power spectrum of the curvature perturbation is
given by
\begin{equation}
\mathcal{P}_{\mathcal{R}}(k) =8\pi\left( \frac {H^2}{2\pi
  \dot{\phi}}\right)^2, \label{power}
\end{equation}
%
and the spectrum of the tensor perturbation by
\begin{equation}
\mathcal{P}_{grav}(k) =\left( \frac {16H^2}{\pi }\right),
\label{gravpower}
\end{equation}
where the right hand sides of these equations are to be evaluated when
the corresponding comoving wavenumber $k$ crosses the horizon. From
this we can also define the tensor-to-scalar ratio
\begin{equation}
r=\frac{\mathcal{P}_{grav}}{\mathcal{P}_{\mathcal{R}}},
\label{r}
\end{equation}
evaluated at some suitable low $k$. Note that there is a factor of
$8\pi$ premultiplying the standard result in equation
(\ref{power}). This is a consequence of the use of different
conventions: $G\equiv 1$ in the standard derivation, whereas here we
have set $8\pi G=\kappa\equiv 1$.

Non-flat universe models, of which Bianchi IX is an example, are more
complicated than their flat counterparts in many respects. However,
they have the decided advantage of having another length scale
available at any time: their associated curvature scale $R$. As set
out below, this allows one to compare length scales at different times
directly, rather than having to use relations from the inferred
history of the universe, such as the epochs of matter and radiation
domination and reheating, the details of which are not well known.
Note that current experiments do indeed allow for curvature
contributions to $\Omega$ at the per cent level, making such
computations {consistent with observations}.

The comoving Hubble radius, and therefore the radius of the spatial
sections, is related to the density parameter by
\begin{equation}\label{closdens}
    \Omega -1=\frac{1}{(RH)^2}.
\end{equation}
The evolution of $\Omega$ for our model is shown in Figure
\ref{figom}.  During inflation, $\Omega$ is driven to unity from
above, whereas after inflation $\Omega=1$ becomes a separatrix rather
than an attractor, and $\Omega$ increases away from unity again.
This provides a relationship linking time, the density
parameter and the curvature radius (and thereby other length scales).
We are particularly interested in when certain length scales cross the
horizon.  The first case of interest is when quantum fluctuations on
different scales leave the horizon during inflation and thereby seed
structure formation. Second, once inflation is completed,
progressively larger scales then re-enter the horizon.  The advantage
of having the curvature scale available in non-flat geometries is that
we can link the two times of horizon crossing for a particular length
scale in a straightforward and exact manner.

Consider some physical size $d_0$ at the present time.  (For wave
modes in such a spherical universe, $d_0$ will actually be quantised
in units of $2\pi R_0/(n^2-1)^{1/2}$, where $n$ is an integer. Here we
will use a continuum approximation, however.)  Scales grow
commensurately with the scale factor during the expansion of the
universe, such that
\begin{equation}\label{growth}
    \frac{d}{R}=\frac{d_0}{R_0}.
\end{equation}
(Here, quantities without indices are evaluated at an arbitrary time,
i.e. their dependence on time is left implicit, whereas the subscript
$0$ denotes quantities evaluated at the present epoch.)  Moreover, the
size of the Hubble radius relative to the radius of the spatial
sections is also changing over the course of cosmic history, as given
by (\ref{closdens}).  Suppose $d_0$ occupies some fraction $x_0$ of
the current Hubble radius $H_0^{-1}$
\begin{equation}\label{physcale}
    x_0=\frac{d_0}{H_0^{-1}}.
\end{equation}
The ratio $x=dH$ will of course in general also change over time and obeys
\begin{equation}\label{xoft}
    x=dH=d_0\frac{R}{R_0}H=x_0\frac{RH}{R_0H_0}.
\end{equation}
Horizon crossing occurs when $x=1$, i.e. the scale with a size $d_0$
at the present time left the horizon (i.e. Hubble length) at a time
$t$ given implicitly by
\begin{equation}\label{horizoncrossing}
    x_0=\frac{R_0H_0}{RH}=\frac{\sqrt{\Omega(t)-1}}{\sqrt{\Omega_0 -1}}.
\end{equation}
where $\Omega_0$ is the present value of the density parameter.  As a
simple example, the scale re-entering the horizon at the moment is, of
course, the current Hubble horizon itself (for which $x_0=1$). Thus,
from (\ref{horizoncrossing}), the present Hubble radius left the
horizon at a time $t$ when $\Omega(t)=\Omega_0$ during inflation. So
for comparison with observations, we are interested in scales that
left the horizon after this epoch.

\condcomment{\boolean{includefigs}}{
\condcomment{\boolean{tikzfigs}}{
\begin{figure}
\tikzstyle{background grid}=[draw, black!50,step=.5cm]
\begin{tikzpicture}
\node (img) [inner sep=0pt,above right]
{\includegraphics[width=8cm]{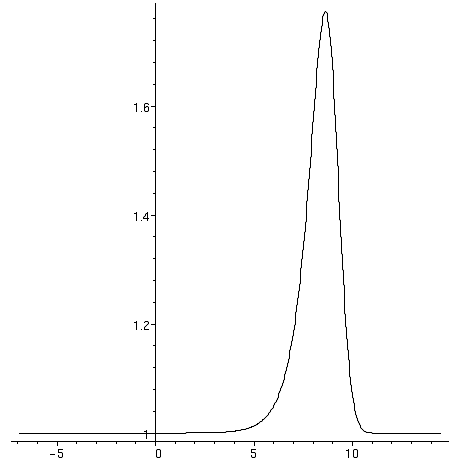}};
\draw (-2:4.5cm) node {$\ln(t)$};
\draw (70:5.5cm) node {$\Omega$};
\end{tikzpicture}
\caption[dummy5]{Evolution of $\Omega$. During inflation the total density is quickly
driven to 1. }
\label{figom}
\end{figure}}

\condcomment{\boolean{psfigs}}{
\begin{figure}
\includegraphics[width=8cm]{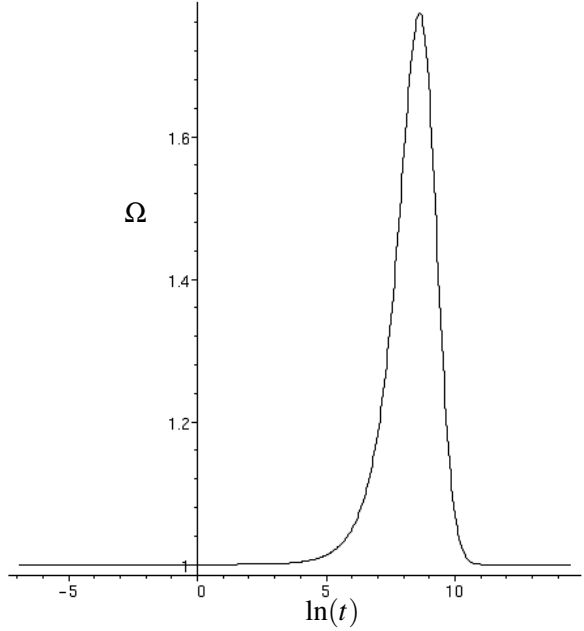}
\caption[dummy5]{Evolution of $\Omega$. During inflation the total density is quickly
driven to 1. }
\label{figom}
\end{figure}}
}

The WMAP 3-year results \cite{Spergel2007WMAP3Cos} quote
$1.011\pm0.012$.  We shall therefore assume that $\Omega_0\sim1.01$.
By the argument above, the time $t$ at which the present Hubble radius
left the horizon is given by the solution of $\Omega(t)=1.01$ during
inflation. This is, by virtue of equation (\ref{closdens}), equivalent to
$-\ln (RH)=-\ln 10 \approx -2.3$, which, referring back to Figure \ref{fig4}, occurs at $\ln
(t)\approx 10.4$.  Comparing with Figure \ref{figpanc}, this corresponds to
about one e-fold of inflation. This still leaves 44 e-foldings
 before the end of inflation, which is sufficient to provide
a perturbation spectrum consistent with experiments, in agreement with
the estimates in \cite{Ellis2003WMAPandCurvature}.  Intriguingly,
however, we can also extract the level of oblateness at that time from
Figure \ref{figpanc} as $\sim 0.2\%$. This is to be regarded as
significantly anisotropic, which means that structure on the largest
scales would have been laid down when the universe was still
oblate. This offers the exciting prospect of a possible experimental
detection of imprints from such a time.

\condcomment{\boolean{includefigs}}{
\condcomment{\boolean{tikzfigs}}{
\begin{figure}
\tikzstyle{background grid}=[draw, black!50,step=.5cm]
\begin{tikzpicture}
\node (img) [inner sep=0pt,above right]
{\includegraphics[width=8cm]{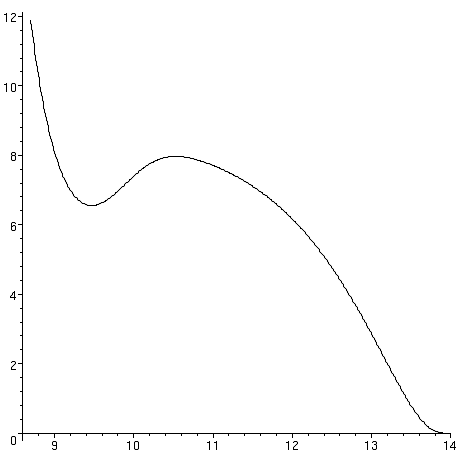}};
\draw (-2:4.5cm) node {$\ln(t)$};
\draw (100:5cm) node {$-10^5 H^2/\dot{\phi}$};
\end{tikzpicture}
\caption[dummy5]{Evolution of $H^2/\dot{\phi}$ during inflation. The function displayed
is $-10^5 H^2/\dot{\phi}$
as a function of $\ln (t/t_p)$. This function determines the amplitude of the curvature perturbation. }
\label{figH2phidot}
\end{figure}}

\condcomment{\boolean{psfigs}}{
\begin{figure}
\includegraphics[width=8cm]{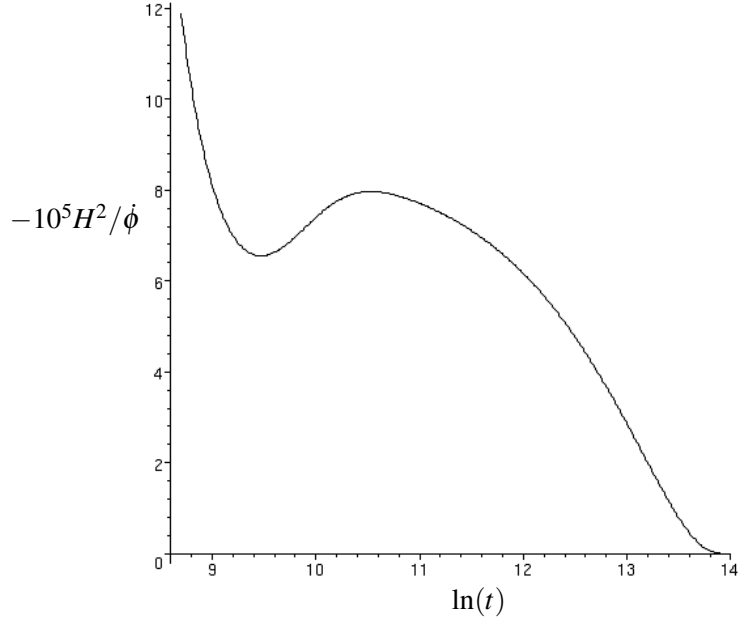}
\caption[dummy5]{Evolution of $H^2/\dot{\phi}$ during inflation. The function displayed
is $-10^5 H^2/\dot{\phi}$
as a function of $\ln (t/t_p)$. This function determines the amplitude of the curvature perturbation. }
\label{figH2phidot}
\end{figure}}

}

Having clarified how to evaluate the quantities given by (\ref{power})
and (\ref{gravpower}), we note that the power spectrum of the
curvature perturbation is controlled by the quantity $H^2/\dot{\phi}$,
plotted in Figure \ref{figH2phidot}. The curvature perturbation
spectrum can now be computed from the joint knowledge of the evolution
of $\Omega$ (Figure \ref{figom}) and $H^2/\dot{\phi}$ (Figure
\ref{figH2phidot}). A physical wavenumber $k_{p,0}$ today is the
inverse of the physical length scale $d_0$ and from (\ref{physcale})
we find the physical wavenumber today that left the horizon at the
time $t$ from (\ref{horizoncrossing}) is given by
\begin{equation}\label{1/k}
    \frac{1}{k_{p,0}}=\frac{x_0}{H_0},
\end{equation}
or, in comoving terms,
\begin{equation}\label{1/kcom}
    \frac{1}{k}=\frac{x_0}{R_0 H_0}.
\end{equation}
Thus for a given comoving wavenumber $k$, equation (\ref{1/kcom}) yields the corresponding value of $x_0$ which
is then substituted into (\ref{horizoncrossing}) to obtain the time $t$ when that scale left the horizon such that
\begin{equation}\label{kOm}
    \frac{1}{k}=x_0\sqrt{\Omega_0 -1}=\sqrt{\Omega(t)-1}.
\end{equation}

\condcomment{\boolean{includefigs}}{
\condcomment{\boolean{tikzfigs}}{
\begin{figure}
\tikzstyle{background grid}=[draw, black!50,step=.5cm]
\begin{tikzpicture}
\node (img) [inner sep=0pt,above right]
{\includegraphics[width=8cm]{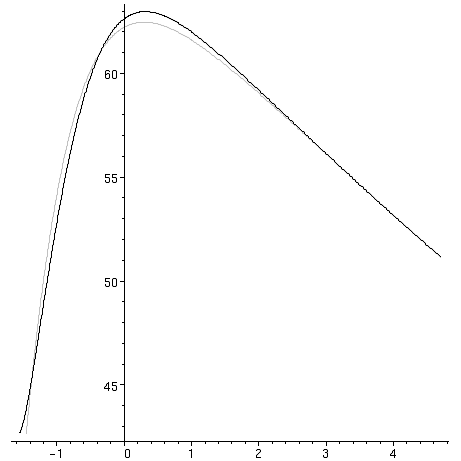}};
\draw (-4:4.5cm) node {$\ln(k)$};
\draw (97:5cm) node {$4\pi^2 10^{7}\mathcal{P}_{\mathcal{R}}(k)$};
\end{tikzpicture}
\caption[dummy5]{The power spectrum of the curvature perturbation. $4\pi^2 10^{7}\mathcal{P}_{\mathcal{R}}(k)$
is plotted as a function of $\ln (1/x_0)=\ln (k) +\constant $. The grey line is a fit to the spectrum with an exponential cutoff proposed by Efstathiou.
 The parameters of the model and the cutoff are described in the text. }
\label{figpert}
\end{figure}}

\condcomment{\boolean{psfigs}}{
\begin{figure}
\includegraphics[width=8cm]{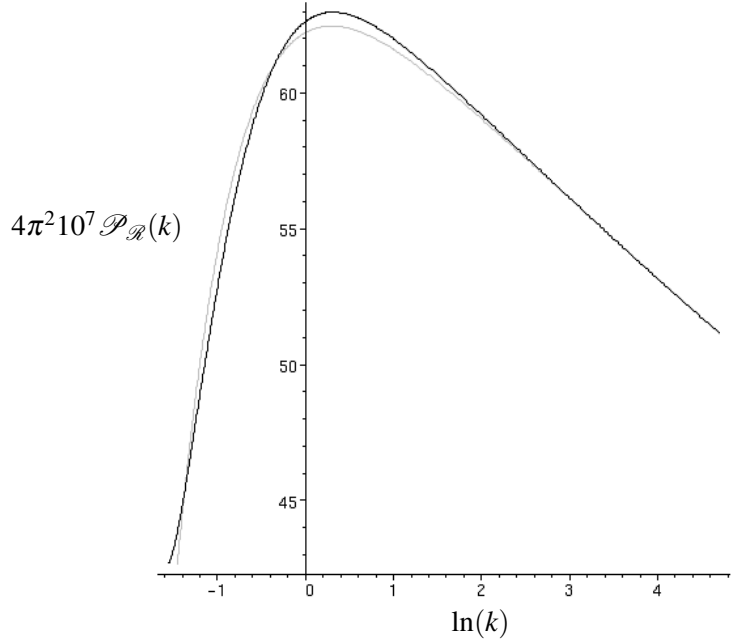}
\caption[dummy5]{The power spectrum of the curvature perturbation. $4\pi^2 10^{7}\mathcal{P}_{\mathcal{R}}(k)$
is plotted as a function of $\ln (1/x_0)=\ln (k) +\constant $. The grey line is a fit to the spectrum with an exponential cutoff proposed by Efstathiou.
 The parameters of the model and the cutoff are described in the text. }
\label{figpert}
\end{figure}}

}

Figure \ref{figpert} shows the curvature power spectrum as a function
of $\ln(1/x_0)=\ln(k)+\constant$ ($=\ln(k_{p,0})+\constant$).
{We again stress that, firstly, due to the isotropisation we are using
an effective FRW-description, such that the direction-dependence  of the modes is negligible to zeroth
order. Secondly, the eigenmodes in a closed universe are discrete -- the spectrum shown here is,
strictly  speaking, a continuous approximation to the underlying discrete spectrum.
It is essentially the direction-averaged power spectrum that is an enveloping function to the
true quantised spectrum.}

The spectrum exhibits a sharp cutoff at low $k$, which is a consequence of the quantity $H^2/\dot{\phi}$ turning over
in the range $t \approx 9.5-10.5$ (which is not the case in conventional flat FRW models with straightforward power laws).
Such an exponential cutoff of the form
\begin{equation}\label{cutoff}
    \mathcal{P}_{\mathcal{R}}(k)=a(1+b\ln(k))(1-\exp(c(\ln(k)+d))
\end{equation}
has in fact been argued for on phenomenological grounds by Efstathiou
\cite{Efstathiou2003Cutoff}. Indeed, our predictions agree rather well
with this proposed exponential cutoff, cf. the grey line in Figure
\ref{figpert}. It is intriguing that our model fits phenomenological
predictions, and, in particular, this could account for the observed
dip at low $\ell$ in the CMB power spectrum. From the power law part
of the spectrum, we can extract the spectral index as $n_s=0.975$,
which is also broadly in agreement with observations
\cite{Spergel2007WMAP3Cos}.  The tensor spectrum is qualitatively very
similar to the scalar spectrum, and yields a tensor-to-scalar ratio of
approximately $r\sim 0.2$ in agreement with current constraints.  Such
agreement with data is encouraging. However, we must again stress that
these computations are to zeroth order and more detailed computations
in the anisotropic setting are needed.

The process of averaging the radii to form
$R=\left(R_1R_2^2\right)^\frac{1}{3}$ is effectively a map from a
deformation of an FRW model back to an FRW model. To zeroth order, we
could then apply all the standard machinery for FRW universes. This is
certainly a very sensible way of achieving an FRW model from a Bianchi
model, but is to some extent arbitrary. Looking at two other FRW
models derived from our Bianchi model is instructive. One could define
an FRW model simply by using each of $R_1$ and $R_2$ in turn.  Of
course, that will lead to problems with singularities at early times,
but for the purpose of calculating fluctuation spectra one could use
the numerical value of either radius in our Bianchi model just before
inflation as an initial condition for an FRW model. It then turns out
that in a model that has the appropriate value of $R_1$ as a starting
point for an FRW universe at the beginning of inflation, the quantity
which controls the magnitude of curvature perturbations,
$H^2/\dot{\phi}$, does not turn over. The curvature spectrum in this
model therefore no longer exhibits an exponential cutoff of the above
mentioned form. A model based on the other radius, $R_2$, however,
does have a turnover similar to the one in Figure \ref{figH2phidot}
and therefore an exponential cutoff in the curvature perturbation
spectrum. This then explains how the unusual feature of an exponential
cutoff arises via the averaging over the two differently behaved
radii.

In summary, the model leads to isotropisation, necessary for
compatibility with standard cosmological models at late times, as well
as inflation, which accounts for structure formation. The perturbation
spectrum we predict meets current constraints on the spectral index
and tensor-to-scalar ratio, and offers an explanation for the dip at
low $\ell$ in the CMB power spectrum. There is also the intriguing
possibility of imprints of early oblateness on structure formation. We
suggest that the universe could have been $0.2$\% oblate at the time
when the present Hubble radius left the horizon.

\section{Perturbations around the Axisymmetric Case \label{pert}}

Given the remarkable properties of this axisymmetric model, it is
important to know how stable it is to perturbations. If a small
fractional perturbation $R_3(t)/R_2(t)=1+\delta(t)$ in the initial
radii evolved to universes in which $R_2(t)$ and $R_3(t)$ were vastly
different, the axisymmetric case would hardly be a viable model for
cosmology. There are several ways in which we can study the stability,
both numerically and analytically.

Setting $R_{3}(t)=R_{2}(t)\left(1+\delta(t)\right)$ in the full
triaxial equations yields the following dynamical equation for the
fractional perturbation $\delta(t)$
\begin{align}
(1+\delta)\ddot{\delta} =  & -\frac{4}{R_1^2}\delta^4-\frac{16}{R_1^2}\delta^3
-\frac{\left(24R_2^2-4R_1^2\right)}{R_1^2R_2^2}\delta^2\notag\\
              &-\left(\left(H_1+2H_2\right)\dot{\delta}+\frac{16R_2^2-8R_1^2}{R_1^2R_2^2}\right)\delta\notag\\
             &-\left(H_1+2H_2\right)\dot{\delta}.
\label{delta}
\end{align}
One can either evolve this $\delta(t)$-equation, or alternatively one
could solve the full triaxial equations
(\ref{generalscalarEoM})-(\ref{H1}) numerically, subject to the
initial fractional perturbation $\delta(0)\ll 1$.

\condcomment{\boolean{includefigs}}{
\condcomment{\boolean{tikzfigs}}{
\begin{figure}
\tikzstyle{background grid}=[draw, black!50,step=.5cm]
\begin{tikzpicture}
\node (img) [inner sep=0pt,above right]
{\includegraphics[width=8cm]{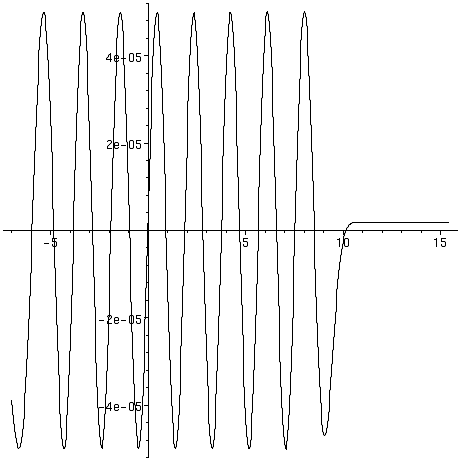}};
\draw (-2:4.5cm) node {$\ln(t)$};
\draw (92:5cm) node {$\ln\frac{R_3}{R_2}$};
\end{tikzpicture}
\caption[dummy5]{Dynamics of the full triaxial Bianchi IX model:
The natural logarithm of the ratio $\frac{R_3}{R_2}$ of the nearly degenerate radii is plotted as a
function of log time (in Planck units). }
\label{figtrix}
\end{figure}}

\condcomment{\boolean{psfigs}}{
\begin{figure}
\includegraphics[width=8cm]{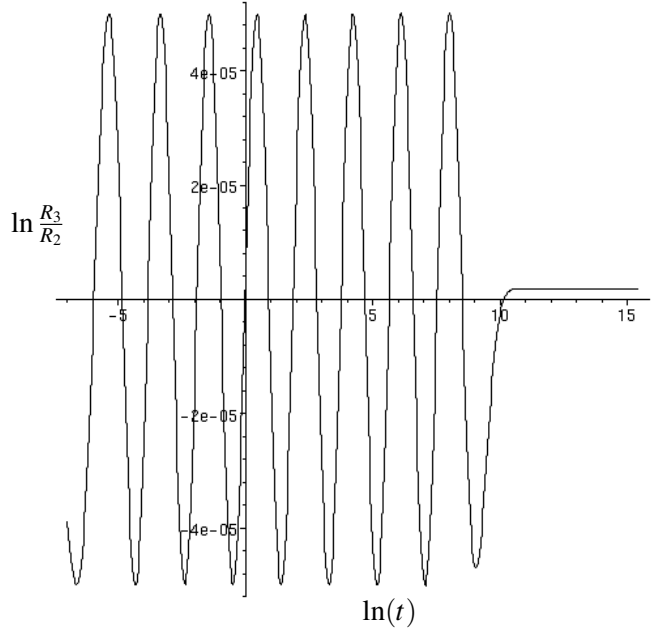}
\caption[dummy5]{Dynamics of the full triaxial Bianchi IX model:
The natural logarithm of the ratio $\frac{R_3}{R_2}$ of the nearly degenerate radii is plotted as a
function of log time (in Planck units). }
\label{figtrix}
\end{figure}}

}

We chose the latter approach, i.e. numerical integration starting from
a point close to the Big Bang with a very small fractional
perturbation $\delta(0)=10^{-4}$ around the biaxial cosmology
considered in the previous section,  keeping the other constants
at their previous values.  The numerical results show that as
$t\rightarrow 0$ ($\ln (t) \rightarrow -\infty$) $R_1$ approaches
zero, but the ratio of $R_3/R_2$ undergoes an infinite set of
oscillations. (The evolution of the quantities corresponding to
Figures \ref{figpanc} to \ref{figom} are indistinguishable from their
biaxial counterparts and hence not shown again.)  This is indeed what
we would expect for the triaxial case (as cited previously,
\cite{RyanShepley1975HomogeneousRelativisticCosmologies}; cf. Figure
\ref{figtrix}). However, we are interested in what happens to these
oscillations at large $t$ rather than at early times.  The results
show that the oscillations cease and the ratio gets frozen in around
the onset of inflation. This shows that for small perturbations a
steady state is eventually reached where the ratio $R_3(t)/R_2(t)$
reaches a constant value. In particular, the resulting value of
$R_3(t)/R_2(t)$ differs from unity by an amount that is similar to the
initial fractional perturbation. Hence the small perturbation does not
lead ultimately to vastly different radii.

Note that here we mean by $t=0$ a different time from in previous
sections. Until now, $t=0$ was naturally defined to be the time where
pancaking occurs. However, perturbing and therefore going to the full
triaxial equations will in general lead to an oscillatory singularity
(see above, \cite{RyanShepley1975HomogeneousRelativisticCosmologies}).
Instead, we now take $t=0$ to mean the point where we impose the
boundary conditions.  These conditions are exactly the same those
applied previously at the Big Bang, except that now the boundary
condition for the perturbed radius $R_3$ is $R_3(0)\equiv
R_2(0)(1+\delta(0))=b_0(1+\delta(0))$.

It is worth noting that, as illustrated in Figure~\ref{figtrix}, the
{\em ratios} of the scale factors become frozen-in.  Thus, even if the
values of the scale factors in each direction differ slightly, the
Hubble functions $H_i(t)$ in each direction will be equal.  This then
raises the question as to whether such a situation would be
observationally distinguishable from an FRW universe at all. It does
not seem to be in the flat case, but the fact that closed and open
universes have an absolute distance scale associated with them via the
curvature scale seems to suggest that it would be detectable. These
considerations will also be described in future work.


In fact the behaviour of the small perturbations around the biaxial case is
reminiscent of the evolution of the conventional curvature perturbations
 generated from quantum fluctuations considered earlier.
 These oscillate on subhorizon scales but get frozen in
on superhorizon scales. They are often regarded as mini-FRW-universes that
are locally over- (or under-) dense and evolve in another FRW-background model.
It seems obvious that these can in general be anisotropic, so that Bianchi
models, in particular Bianchi IX, might be good for describing cosmological perturbations even when the
background model is strictly Friedmann-Robertson-Walker. The different behaviour
on subhorizon and superhorizon scales is normally explained by saying that
subhorizon scales describe causally connected regions where `there is time
for local differences in matter distribution to affect the physics'.
Superhorizon scales, however, are causally disconnected and frozen in.
The cosmologically relevant scales that were previously subhorizon are stretched
to superhorizon scales by inflation, and are now slowly coming back into the horizon.
Therefore it is only natural to think of the perturbations in the Bianchi
model in the same way -- as stretched to superhorizon scales by inflation
and frozen in.

We can make this empirical connection slightly more quantitative.
Consider the evolution of a curvature perturbation associated with a
comoving wavenumber $k$ in a flat FRW-universe
\cite{HobsonLasenbyEfstathiou2006GeneralRelativity}. (Note that this
is an excellent approximation for our model after the start of
inflation.)  This is given by
\begin{equation}\label{HLE}
    \ddot{\zeta}_k+\left( \frac{\dot{\phi}^2}{H} +2 \frac{\ddot{\phi}}{\dot{\phi}} +3H\right)
    \dot{\zeta}_k+\frac{k^2}{R^2}\zeta_k=0.
\end{equation}
This equation is of harmonic oscillator form. However, amplitude and
frequency are time-dependent.  Thus it is reasonable that, in general,
the motion will be oscillatory. At late times (after inflation),
however, the effects of the scalar field driving inflation are
negligible and the Hubble radius is large, so overall the damping term
is negligible. Also the size of the universe is then large, so that
the equation basically reduces to $\ddot{\zeta}_k=0$. Thus at late
times the perturbation gets frozen in.

Now consider perturbations around the biaxial case (\ref{delta}) in
the late-time, isotropic limit $H_i=H$, $R_i=R$. This yields
\begin{equation}\label{deltalate}
    \ddot{\delta}+3H\dot{\delta}+\frac{8}{R^2}\delta=0.
\end{equation}
Thus the two different kinds of perturbation actually obey very
similar equations. They are practically of the same form when the
effect of the scalar field is negligible (after inflation).  The fact
that the terms involving the respective function and its first
derivative are suppressed by the expansion of the universe explains
why both types of oscillation become frozen-in eventually.  From the
equations, we would, however, expect the evolution of $\delta$ to
exhibit much less variation of amplitude and period with time, which
is indeed what is observed numerically.  Hence, at the level of the
dynamical equations, it is reasonable that the two different sorts of
perturbation look so similar. The deeper question of why they should
obey such similar equations will be considered in a future work.


One could, alternatively, argue that the more natural variable to
consider is the fractional perturbation in the Hubble functions,
rather than the radii.  We choose to parameterise this fractional
perturbation in an analogous way as $H_3(t)=H_2(t)(1+h(t))$. Its
evolution is governed by the equation
\begin{align}\label{h}
    \dot{h}= & -\frac{1}{2}(H_1+H_2)h^2\notag\\
    & -\frac{1}{2}\left(2H_1H_2+H_2^2+3\frac{R_1^2}{R_2^2R_3^2}
    +3\frac{R_3^2}{R_1^2R_2^2}-5\frac{R_2^2}{R_1^2R_3^2}+2\frac{1}{R_1^2}+2\frac{1}{R_3^2}
    -6\frac{1}{R_2^2}-\kappa p+\Lambda\right)\frac{h}{H_2}\notag\\
    & -4\left(\frac{R_3^2}{R_1^2R_2^2}
    -\frac{R_2^2}{R_1^2R_3^2}+\frac{1}{R_3^2}-\frac{1}{R_2^2}\right)\frac{1}{H_2}.
\end{align}
From this equation it can be seen that, in fact, an expansion in which
two axes have the same Hubble function is favoured and stable: for a
small initial fractional perturbation the terms in powers of $h$ are
negligible, and the constant term vanishes when $R_2=R_3$, which
implies that $h=0$ and, from (\ref{h}), $\dot{h}=0$. This further
supports the claim that this biaxial model is stable and hence a
viable and interesting cosmological model.

\section{Bounce solution\label{bounce}}
In addition to the odd-parity{, pancaking} solution that we have discussed so far,
there also exists an even-parity, bouncing solution to the biaxial
evolution equations (\ref{special1}--\ref{specialEoM}).  In this
solution, all radii are finite at the Big Bang and go smoothly through
into a pre-Big Bang phase. The series solution ansatz akin to
(\ref{seriesansatz}) is {even under parity}
\begin{align}
R_{1}(t)&=a_{0}+a_{2}t^{2}+a_{4}t^{4}+\dots\notag\\
R_{2}(t)=R_{3}(t)&=b_{0}+b_{2}t^{2}+b_{4}t^{4}+\dots \,\,\,\,. \notag\\
\phi(t)&=f_{0}+f_{2}t^{2}+f_{4}t^{4}+\dots \notag
\\\label{seriesansatzeven}
\end{align}
In contrast to the previous series solution, the Friedmann constraint
is not automatically satisfied, and leads to an additional constraint
determining $a_0$ in terms of $b_0$, $f_0$, etc.  Figure
\ref{figbounce} shows the basic dynamics of this model. Both radii are
constant and non-zero at the Big Bang, and the effect of the scalar
field is again to inflate and isotropise the universe.
{For considerations concerning a presentation in terms of shear and curvature,
see the appendix.}
This model
might be interesting in its own right, and further details will be
presented elsewhere. {For previous work on bouncing solutions
and their likelihood see, for example, \cite{Barrow1980SizeOfBouncingMixmaster, Lilley2008Bounce}.}

\condcomment{\boolean{includefigs}}{
\condcomment{\boolean{tikzfigs}}{
\begin{figure}
\tikzstyle{background grid}=[draw, black!50,step=.5cm]
\begin{tikzpicture}
\node (img) [inner sep=0pt,above right]
{\includegraphics[width=8cm]{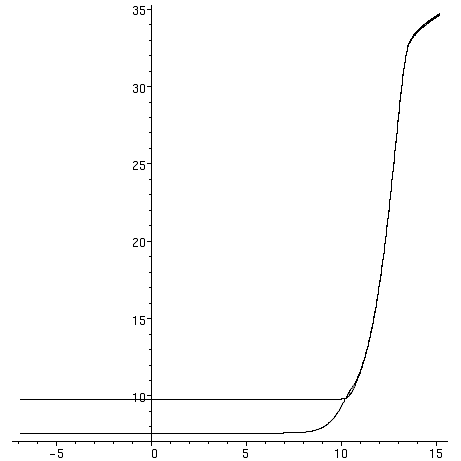}};
\draw (-5:5.5cm) node {$\ln(t)$};
\draw (115:1.2cm) node {$\ln\left(R_2(t)\right)$};
\draw (132:0.8cm) node {$ \ln\left(R_1(t)\right)$};
\end{tikzpicture}
\caption[dummy1]{Dynamics of the bouncing solution:
 evolution of the logarithm of the scale factors $R_1$ and $R_2$ in Planck
lengths $l_p$ versus log time ($t$ in units of Planck time $t_p$).}
\label{figbounce}
\end{figure}}

\condcomment{\boolean{psfigs}}{
\begin{figure}
\includegraphics[width=8cm]{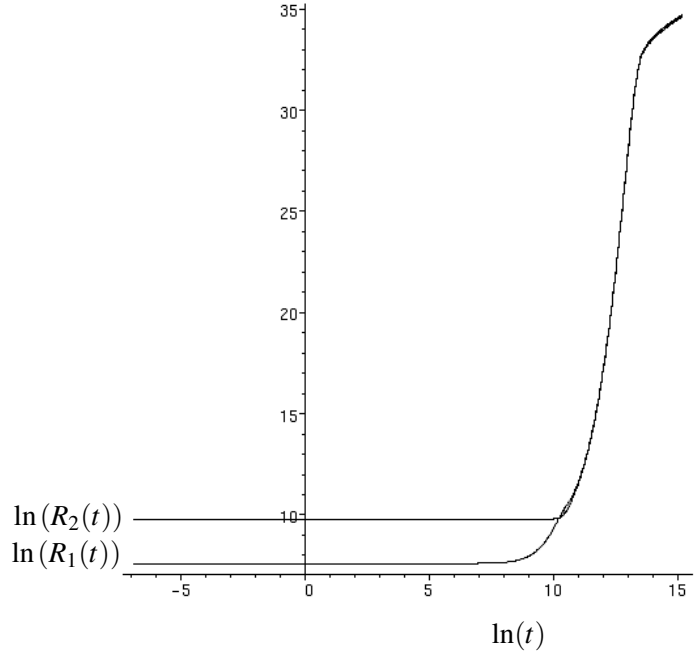}
\caption[dummy1]{Dynamics of the bouncing solution:
 evolution of the logarithm of the scale factors $R_1$ and $R_2$ in Planck
lengths $l_p$ versus log time ($t$ in units of Planck time $t_p$).}
\label{figbounce}
\end{figure}}

}

\section{Conclusions\label{con}}

We have presented a novel scenario in which the effect of a scalar
field and biaxial Bianchi IX geometry is to render the Big Bang much
better behaved.  Physical quantities remain finite at the Big Bang,
and there is no curvature singularity at the time of pancaking. This
is in contrast with singularities in flat and open FRW-models, and
essential singularities in the full triaxial Bianchi IX case.  In
addition, the investigation of the behaviour of geodesics suggests
that some physical observers can cross over and move smoothly into a
pre-Big Bang phase. Some observers, however, may not be able to cross
and this would appear to make the model incomplete, with the pancake
being a quasiregular singularity (the type that also occurs in
Taub-NUT). However, in the light of periodic identification of angular
coordinates on the 3-sphere and the fact that the winding direction is
the collapsing dimension, this might not be such a problem.  The model
also exhibits stability under sufficiently small perturbations around
biaxiality.  Though the model isotropises at late times, we show that
structure on the largest scales could in fact stem from a time at
which the universe was significantly anisotropic.  Such behaviour is
very desirable for a satisfactory cosmological model, as a closed
universe can be matched to the observed asymptotic de Sitter phase of
$\Lambda$- domination, whilst the scalar field can drive
isotropisation, so as to account for the observed degree of isotropy
of the universe, and inflation, thought to be needed in order to
explain the observed perturbation spectrum. We predict a spectral
index and a tensor-to-scalar ratio compatible with current
constraints, as well as a dip at low multipoles of the CMB power
spectrum, which is consistent with an exponential cutoff that has also
been argued for on phenomenological grounds.  We explain the apparent
qualitative similarity between the two very different kinds of
perturbations, namely curvature perturbations and perturbations around
biaxiality, from their evolution equations.  A separate, bouncing
solution is also presented, and will be considered in more detail
elsewhere. A more thorough analysis of many subtleties of the model,
in particular the issue of geodesic completeness and a detailed
comparison with Taub-NUT is in preparation.

\begin{acknowledgements}
{We  thank the  referee for their very useful suggestions.
We also thank John Barrow, Sylvain Br\'echet, Carlo Contaldi,
Leonardo Fern\'andez-Jambrina,
Cyril Pitrou, Paolo Salucci,  Subir Sarkar, Jiro Soda and others
for helpful comments and for pointing out relevant references.
PPD is grateful for support through an STFC (formerly PPARC)
studentship. }
\end{acknowledgements}

\appendix*
\section{Dynamics in the $3+1$ covariant approach\label{app}}

{Here we use the  timelike eigenvector $u_\mu=[1,0,0,0]$ of the scalar field energy-momentum tensor
as the `velocity field' to perform a $3+1$ split
(see, for example, \cite{Ellis1998Cargese})
in order to present the equations (\ref{special1})-(\ref{specialEoM}) in a
different parametrisation, which might be convenient for other applications, and in particular
separates more clearly the contributions of curvature and shear. Many quantities in the general $3+1$ approach
vanish for our particular case, such as the peculiar acceleration, vorticity and anisotropic stress.
 The dynamical equations can therefore be recast simply in terms of the averaged scale factor, i.e. the volume expansion,
\begin{equation}
R\equiv \left(R_1R_2R_3\right)^{\frac{1}{3}}\rightarrow\left(R_1R_2^2\right)^{\frac{1}{3}},
\label{spvolumeexp}
\end{equation}
its associated Hubble factor,
\begin{equation}
3H= H_1+H_2+H_3\rightarrow H_1+2H_2,
\label{spvolumeexpH}
\end{equation}
and the shear scalar $\sigma^2$
\begin{equation}
\sigma^2=\frac{1}{6}\left[\left( H_1-H_2\right)^2+\left( H_2-H_3\right)^2+\left( H_3-H_1\right)^2\right]\rightarrow
\frac{1}{3}\left( H_1-H_2\right)^2,
\label{spshearscalar}
\end{equation}
where the expressions on the right denote the biaxial limit.
}
{The shear tensor $\sigma_{\mu\nu}$ is defined as
$\sigma_{\mu\nu}\equiv \tilde{\nabla}_{\langle\mu}u_{\nu\rangle}$,
where  $\tilde{\nabla}$ is the fully orthogonally projected covariant derivative
(for details, see \cite{Ellis1998Cargese}).
The shear tensor has three spacelike eigenvectors characterised in a coordinate-free way by their eigenvalues.
In the biaxial case, for instance, one is 
$\frac{2}{3}\left( H_1-H_2\right)$, whereas the other two are degenerate and have the value
$-\frac{1}{3}\left( H_1-H_2\right)$, whence the above shear scalar follows from
$\sigma^2\equiv \frac{1}{2}\sigma_{\mu\nu}\sigma^{\mu\nu}.$
}

In this parametrisation, the Raychudhuri equation assumes the form
\begin{equation}
3\dot{H}+3H^2=-\frac{\kappa}{2}\left(\rho+3p\right)-2\sigma^2+\Lambda.
\label{altRaychaudhuri}
\end{equation}
The Friedmann equation can be reexpressed as
\begin{equation}
H^2=\frac{\kappa\rho}{3}+\frac{\sigma^2}{3}+\frac{\Lambda}{3}+\frac{R^{(3)}}{6},
\label{altFriedmann}
\end{equation}
where $R^{(3)}$ is the Ricci scalar of the orthogonal 3-spaces, which here evaluates to
\begin{equation}
R^{(3)}=2\left(    \frac{R_1^2}{R_3^2 R_2^2}+\frac{R_2^2}{R_3^2 R_1^2}+\frac{R_3^2}{R_1^2 R_2^2}
-\frac{2}{R_1^2}-\frac {2}{R_2^2}-\frac {2}{R_3^2}   \right)\rightarrow \frac{2}{R_2^2}\left(\frac{R_1^2}{R_2^2}-4\right).
\label{altRicci}
\end{equation}
This uncovers the geometric significance of the corresponding term in the triaxial Friedmann equation (\ref{Fr}).
We also find that the $-H_1H_2-H_2H_3-H_3H_1$ term just corresponds to $3H^2-\sigma^2$ (up to sign),
and $3H^2+2\sigma^2=H_1^2+H_2^2+H_3^2$.
This also correctly reduces to the standard FRW results in the isotropic limit, as do the equations themselves.
The usual shear propagation for vanishing peculiar acceleration and anisotropic stress is  %
\begin{equation}
\dot{\sigma}_{\langle\mu\nu\rangle}+3H\sigma_{\mu\nu}=R^{(3)}_{\mu\nu}-\frac{1}{3}h_{\mu\nu}R^{(3)},
\label{genshearprop}
\end{equation}
where $R^{(3)}_{\mu\nu}$ and $R^{(3)}$ are the Ricci tensor and scalar of the orthogonal 3-spaces, respectively.
However, in the biaxial case, the evolution equation for the shear scalar suffices, and turns out to be
\begin{equation}
\dot{\sigma}+3H\sigma=\frac{1}{2\sigma}R^{(3)}_{\mu\nu}\sigma^{\mu\nu}=-\frac{4}{\sqrt{3}R_2^2}\left(\frac{R_1^2}{R_2^2}-1\right).
\label{altshear}
\end{equation}
This can be seen upon contracting (\ref{genshearprop}) with $\sigma^{\mu\nu}$ and noting that
$\dot{\sigma}_{\langle\mu\nu\rangle}\sigma^{\mu\nu}=\dot{\sigma}_{\mu\nu}\sigma^{\mu\nu}
=\frac{1}{2}({\sigma}_{\mu\nu}\sigma^{\mu\nu})^\cdot=2\sigma\dot{\sigma},$
i.e. since one is contracting with something purely spatial, the projection onto
the 3-spaces prior to the contraction does not change the result.
Furthermore, the term involving the Ricci scalar of the 3-surfaces just extracts the trace of the shear, which
vanishes by definition.

The equation of motion for the scalar field just retains its usual Klein-Gordon form
\begin{equation}
  m^{2}\phi+3H\dot{\phi}+\ddot{\phi}=0.
\label{altspecialEoM}
\end{equation}

{Using the series solution (\ref{seriesansatz}) for the pancaking solution, we find that
\begin{equation}
 R(t)=\left(R_1R_2^2\right)^{\frac{1}{3}}\propto t^{\frac{1}{3}},
\label{pancR}
\end{equation}
to lowest order in $t$. By the same token, the shear scalar behaves as
\begin{equation}
\sigma^2=\frac{1}{3}\left( \frac{R_2}{R_1}\right)^2\left(\frac{d}{dt}\left(\frac{R_1}{R_2}\right)\right)^2
=\frac{1}{3}t^{-2}+\textit{O}(1)\propto R^{-6}+{\textit{O}}(1),
\label{pancshear}
\end{equation}
as might be expected from the naive form of equation (\ref{altshear}) $\dot{\sigma}+3H\sigma=0$, which holds in the
better-known flat and isotropic limits. Note that this
 does not contradict equation (\ref{altshear}). The lowest order term in the series expansion for $\sigma$ is
 $\sigma\propto H_1(t)\propto \frac{1}{t}$ which results in
 $\dot{\sigma}\propto-\frac{1}{t^2}$ to lowest order. This happens to cancel exactly the $\frac{1}{t^2}$ term
 resulting from the product of $\sigma\propto \frac{1}{t}$ and $H(t) \propto H_1(t)\propto \frac{1}{t}$, such that the
 leading order of the combined  $\dot{\sigma}+3H\sigma$ is actually the constant that is the leading term
 on the right hand side of (\ref{altshear}).
 Also note that the approximate time-dependence of the shear scalar does not
involve any of the parameters that need to be set in our model, which is remarkable.
The curvature of the orthogonal 3-space is constant across the pancake
\begin{equation}
R^{(3)}=-\frac{8}{b_0^2},
\label{pancR3}
\end{equation}
whereas the shear diverges, as shown. This means that in a pancaking Bianchi IX model
the shear will always dominate over the curvature at early enough times, such that the model is
actually well approximated by a Bianchi I model, though they are of course different topologically.
}
\condcomment{\boolean{includefigs}}{
\condcomment{\boolean{tikzfigs}}{
\begin{figure}
\tikzstyle{background grid}=[draw, black!50,step=.5cm]
\begin{tikzpicture}
\node (img) [inner sep=0pt,above right]
{\includegraphics[width=8cm]{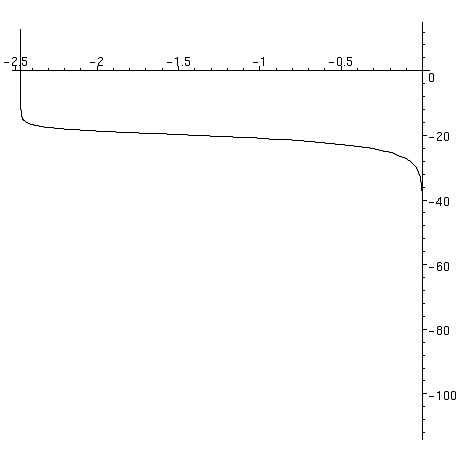}};
\draw (95:7.2cm) node {$10^8R^{(3)}$};
\draw (48:11.3cm) node {$ \ln\sigma$};
\draw (1.9,8.52) node {Pancake};
\draw (5.8,3.15) node {FRW};
\draw[latex-] (0.4,7.52) .. controls +(.5,0) and +(-.5,-.5) .. +(1.5,0.5);
\draw[latex-] (7.3,4.4) .. controls +(0,0) and +(0,0) .. +(-1.5,-1);
\draw[latex-] (5,5.2)  -- +(-4,0.15);
\end{tikzpicture}
\caption[dummy1]{$\left(\sigma,R^{(3)}\right)$-trajectory for the pancaking solution:
 evolution of the logarithm of the shear versus the Ricci scalar of the orthogonal 3-surfaces.}
\label{figpancphasespace}
\end{figure}}

\condcomment{\boolean{psfigs}}{
\begin{figure}
\includegraphics[width=8cm]{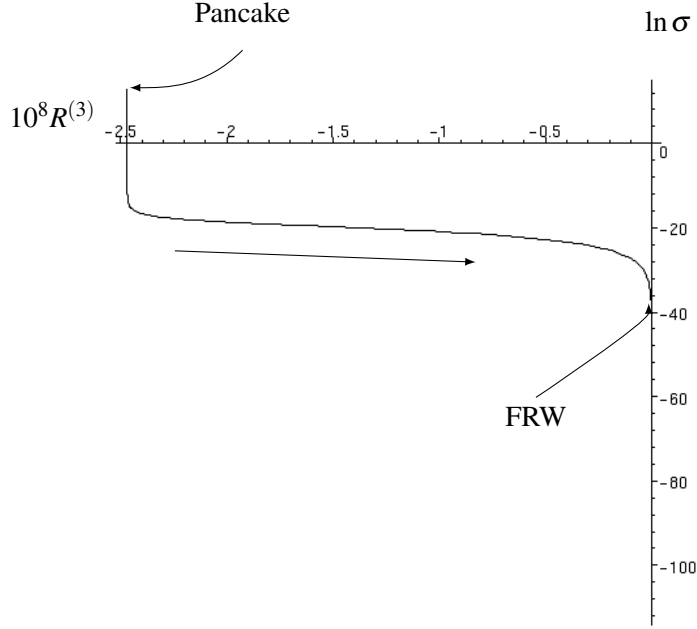}
\caption[dummy1]{$\left(\sigma,R^{(3)}\right)$-trajectory for the pancaking solution:
 evolution of the logarithm of the shear versus the Ricci scalar of the orthogonal 3-surfaces.}
\label{figpancphasespace}
\end{figure}}

}
{Figure \ref{figpancphasespace} depicts the motion of the pancaking model through
the shear-curvature-parameter space. Note that since the shear diverges at the time of pancaking
we have chosen to plot the model's trajectory in the $\left(\ln \sigma, R^{(3)}\right)$-space  rather than the
$\left(\sigma, R^{(3)}\right)$-space  itself (which will be appropriate for the bouncing solution).
At early times, the shear diverges, whereas the curvature is constant, which accounts for the divergence
in the top left of the plot corresponding to the time of pancaking. During inflation the curvature is quickly driven to zero and the
universe isotropises i.e. the shear vanishes, which accounts for the late-time behaviour in the bottom right part. }

{Similarly,  for the bouncing solution (\ref{seriesansatzeven}), we find that
\begin{equation}
 R(t)\propto \constant,
\label{bouncR}
\end{equation}
and also the shear is constant (in fact vanishes) to first order
\begin{equation}
\sigma^2=\frac{4}{3}\left( \frac{a_2}{a_0}-\frac{b_2}{b_0}\right)^2 t^2\approx 0,
\label{bouncshear}
\end{equation}
as might be expected, since all the dynamical variables have an extremum at the bounce.
The Ricci scalar of the 3-surfaces is also constant
\begin{equation}
R^{(3)}=\frac{2}{b_0^2}\left(\frac{a_0^2}{b_0^2}-4\right).
\label{bouncR3}
\end{equation}
This means that for a bouncing model, contrary to the pancaking case,
the curvature will generically dominate over
the shear at the bounce.
}

\condcomment{\boolean{includefigs}}{
\condcomment{\boolean{tikzfigs}}{
\begin{figure}
\tikzstyle{background grid}=[draw, black!50,step=.5cm]
\begin{tikzpicture}
\node (img) [inner sep=0pt,above right]
{\includegraphics[width=8cm]{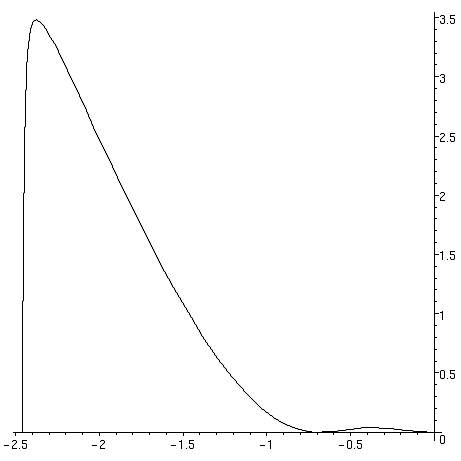}};
\draw (48:11.3cm) node {$10^9\sigma$};
\draw (145:0.7cm) node {$10^8R^{(3)}$};
\draw (2,2) node {Bounce};
\draw (5.8,1.8) node {FRW};
\draw[latex-] (0.4,0.52) .. controls +(.5,0) and +(-.5,-.5) .. +(1.5,1);
\draw[latex-] (7.55,0.5) -- +(-1.5,1);
\draw[latex-] (2,6.2) .. controls +(-2,4) and +(0,0) .. +(-2,0);
\end{tikzpicture}
\caption[dummy1]{$\left(\sigma,R^{(3)}\right)$-trajectory for the bouncing solution:
evolution of the shear versus the Ricci scalar of the orthogonal 3-surfaces.}
\label{figbouncephasespace}
\end{figure}}

\condcomment{\boolean{psfigs}}{
\begin{figure}
\includegraphics[width=8cm]{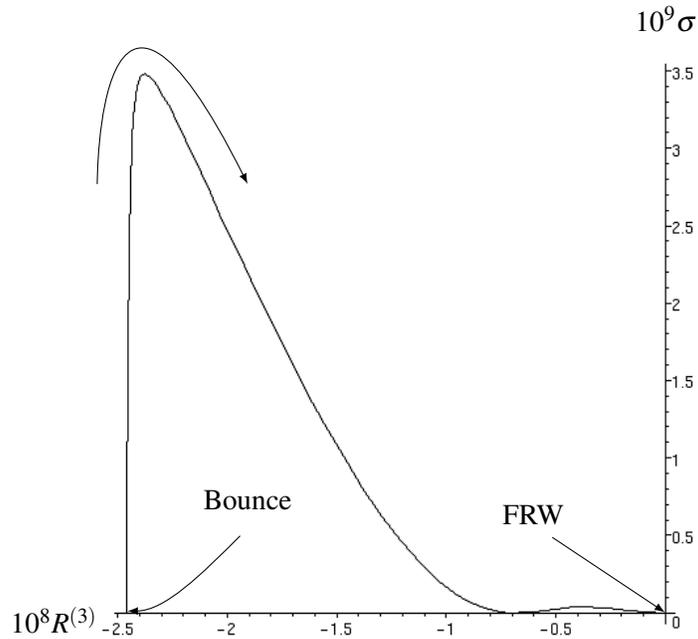}
\caption[dummy1]{$\left(\sigma,R^{(3)}\right)$-trajectory for the bouncing solution:
evolution of the shear versus the Ricci scalar of the orthogonal 3-surfaces.}
\label{figbouncephasespace}
\end{figure}}

{Figure \ref{figbouncephasespace} shows the motion of the bouncing model through
the shear-curvature-parameter space.
At early times, the shear vanishes, and likewise at late times. The shear therefore peaks
at intermediate times in the bouncing model. The curvature shows the same behaviour as the pancaking
solution in that it is constant at early times and the model is then driven to flatness during inflation.
So we can identify the bottom left hand area of the plot  with the bounce, where the curvature is maximal and the
shear vanishes. The shear then peaks at intermediate times whilst the curvature
decreases. The bottom right hand corner again corresponds to the late-time evolution which is
driven to isotropy and spatial flatness. Note that in both cases the numerical value of the
initial curvature is exactly what we expect for the chosen parameter values. They appear to be nearly
the same  since for the particular models chosen here  $a_0\ll b_0$. }

}

\bibliography{bianchibib2}

\end{document}